\documentclass{aa}  

\usepackage{graphicx}
\usepackage{txfonts}
\usepackage{amsmath}
\usepackage{amssymb}	
\usepackage{commath}    
\usepackage{tabulary}   
\usepackage{tabularx}   
\usepackage{array}
\usepackage{threeparttable}  
\usepackage{booktabs}   
\usepackage[colorlinks,citecolor=blue]{hyperref}
\graphicspath{{figs/}}

\makeatletter
\newcommand\thefontsize{\f@size pt}
\makeatother
\newcommand{\codename}[2][]{\mbox{\textsc{#2}}#1 }

\newcommand{\ngjfid}{n_{\rm GJ}}
\newcommand{\rh}{r_H}

\newcommand{\dfid}{d_{e0}}
\newcommand{\multiplicity}{\eta}
\newcommand{\ombh}{\Omega_{\rm BH}}
\newcommand{\btoyfid}{B_{r_0}}
\newcommand{\dissprofvar}{\dot{\epsilon}}
\newcommand{\bundleidx}{b}

\usepackage[normalem]{ulem}
\begin{document}

   \title{A kinetic model of jet-corona coupling in accreting black holes}

   \author{J.\ Mehlhaff\inst{1}
          \and
          B.\ Cerutti\inst{1}
          \and
          B.\ Crinquand\inst{2,3}
          }

         \institute{Univ.\ Grenoble Alpes, CNRS, IPAG,
              38000 Grenoble, France\\
              \email{john.mehlhaff@univ-grenoble-alpes.fr}
              \and
              IRAP, Université de Toulouse, CNRS, CNES, UPS,
              14 avenue Edouard Belin, 31400 Toulouse, France
              \and
              Department of Astrophysical Sciences, Peyton Hall,
              Princeton University, Princeton, NJ 08544, USA
             }

   \date{Received \today; accepted --}

  \abstract
   {
       Black hole (BH) accretion disks are often coupled to ultramagnetized and tenuous plasma coronae close to their central BHs. The coronal magnetic field can exchange energy between the disk and the BH, power X-ray emission, and lead to jetted outflows.
       Up until now, the coronal physics of BH accretion has only been studied using fluid modeling.
   }
   {
       We construct the first model of a BH feeding on a zero-net-flux accretion disk corona based on kinetic plasma physics. This allows us to self-consistently capture how collisionless relativistic magnetic reconnection regulates the coronal dynamics.
   }
   {
       We present global, axisymmetric, general relativistic particle-in-cell simulation of a BH coupled, via a series of magnetic loops, to a razor-thin accretion disk.
       We target the jet-launching regime where the loops are much larger than the BH.
       We ray-trace high-energy synchrotron lightcurves and track the flow of Poynting flux through the system, including along specific field-line bundles.
   }
   {
       Reconnection on field lines coupling the BH to the disk dominates the synchrotron output, regulates the flux threading the BH, and ultimately untethers magnetic loops from the disk, ejecting them via a magnetically striped Blandford-Znajek jet. 
       The jet is initially Poynting-dominated, but reconnection operates at all radii, depleting the Poynting power logarithmically in radius.  
   }
   {
       Coronal emission and jet launch are linked through reconnection in our model. This link might explain coincident X-ray flaring and radio-jet ejections observed during hard-to-soft X-ray binary state transitions.
       It also suggests that striped jet launch could be heralded by a bright coronal counterpart.
       Our synchrotron signatures resemble variability observed from the peculiar changing-look AGN,~1ES~1927+654, and from Sagittarius A*, hinting that processes similar to our model may be at work in these contexts.
   }

   \keywords{Acceleration of particles --
            Accretion, accretion disks --
            Black hole physics --
            Magnetic reconnection --
            Plasmas
    }
   \maketitle

\nolinenumbers
\section{Introduction}
\label{sec:intro}
Soon after the advent of the standard theory of accretion disks \citep{shakura_sunyaev_1973, novikov_thorne_1973}, it was recognized \citep[][]{bisnovatyiKogan_blinnikov_1976, liang_price_1977} that a classic geometrically thin, optically thick accretion disk cannot supply the hard X-ray radiation (up to~$100 \, \rm keV$ or so) observed in black hole (BH) X-ray binaries (XRBs). This led to the postulate that such disks may exist in close proximity to a much hotter and more tenuous plasma. The putative plasma was dubbed, by analogy with the hot and tenuous X-ray emitting plasma enshrouding the Sun, an accretion disk corona \citep[ADC;][]{bisnovatyiKogan_blinnikov_1976, liang_price_1977}.

Pursuing the solar analogy farther, \citet{galeev_etal_1979} suggested that convective instabilities in the disk generate magnetic fields that subsequently buoyantly rise up out of the disk, leading to an arcade of ultramagnetized magnetic loops tethered to, and manipulated by, the heavy accretion disk below \citep[see also][]{tout_pringle_1992}. In this scenario, magnetic fields are not only the hallmark of the coronal morphology, but also constitute the main vehicle through which free energy from the underlying disk rotation can be transformed into hard X-ray emission. For example, relativistic magnetic reconnection \citep{blackman_field_1994, lyutikov_uzdensky_2003, lyubarsky_2005}, can efficiently transfer magnetic energy to the plasma, offsetting the extremely rapid radiative losses (\citealt{galeev_etal_1979, diMatteo_1998, liu_etal_2002, deGouveiaDalPino_lazarian_2005, uzdensky_goodman_2008, goodman_uzdensky_2008, singh_etal_2015, kadowaki_etal_2015, khiali_etal_2015, uzdensky_2016, beloborodov_2017}; though magnetized turbulence may be a complementary driver of particle energization, \citealt{singh_etal_2015, kadowaki_etal_2015, groselj_etal_2024, nattila_2024}).

Besides merely \textit{dissipating} magnetic energy, reconnection also \textit{reconfigures} the magnetic field lines, thereby modifying how they transport energy and angular momentum through the corona. Powering high-energy radiation is thus not the only important role reconnection may play in ADCs. It may also profoundly impact global aspects of accretion, including concomitant ejection phenomena.

For example, reconnection may compete with Keplerian shear and random magnetic footpoint motion to set the coronal scale height and regulate the balance between closed and open field-line bundles \citep{tout_pringle_1996, uzdensky_goodman_2008}. By shaping the closed portion of the corona, reconnection sets the efficiency with which angular momentum is transported from inner to outer closed-loop footpoints, directly impacting accretion \citep{heyvaerts_priest_1989, goodman_2003}. At the same time, the influence of reconnection on open field lines is equally important, since these can channel a magneto-centrifugal wind \citep{blandford_payne_1982} and thus directly shed angular momentum off of the disk \citep{konigl_1989}.

Furthermore, in the event that a coronal magnetic link is established between the accretion disk and a spinning central BH, magnetic reconnection can throttle the magnetic flux threading the event horizon, thereby controlling the extraction of BH spin energy in the form of a \citet{blandford_znajek_1977} jet. If such a coupling is repeated, with multiple opposing-polarity coronal loops successively fed to the BH and ejected via the Blandford-Znajek (BZ) mechanism \citep{parfrey_etal_2015}, the nascent jet is embedded with a striped magnetic configuration that may dissipate much farther downstream \citep{giannios_uzdensky_2019}. Clearly, then, by shaping ADCs, reconnection is intricately tied to the more general phenomena of BH accretion, outflow, energy extraction, and jet dissipation.

These mainly theoretical remarks are underscored by observations, especially those of outbursting BH XRB state transitions. In a summary of the main observed features of such outbursts, \citet{fender_etal_2004} pointed out that the binary's transition to a softer X-ray spectrum and reduced X-ray variability is generally hailed by one or more discrete radio-jet ejections (see also \citealt{fender_etal_2009}, \citealt{miller-jones_etal_2012}, and \citealt{russell_etal_2019} as well as the reviews by \citealt{remillard_mcclintock_2006}, \citealt{done_etal_2007}, and \citealt{ingram_motta_2019}). This bolsters the expected picture in which reconnection-mediated coronal structure, reconnection-powered X-ray flaring, and jet launch are all connected, as indeed has already been appreciated in the literature. \citet{tagger_etal_2004} and \citet{deGouveiaDalPino_lazarian_2005}, for example, posited that the radio-jet ejections are created by reconnection in the inner corona along field lines coupling the accretion disk to the central BH.

Modeling ADCs requires, as a first step, choosing a suitable mathematical description of the coronal plasma. Here, the framework of force-free electrodynamics is often employed, since, barring thin reconnecting current sheets, the corona is thought to be well magnetized so that the plasma inertia and the bulk electromagnetic force both approximately vanish. As a result, scenarios in which coronal activity above razor-thin accretion disks links to jet launch have chiefly been studied in the force-free framework \citep{parfrey_etal_2015, yuan_etal_2019, mahlmann_etal_2020}. These works have shown that magnetically coupling a spinning BH to its accretion disk can, for the right coronal loop structure, permit strong BZ jet production. The basic constraint is that the loops must be much larger than the BH. Otherwise, rotational shear is insufficient to open the field lines threading the BH horizon \citep{uzdensky_2005, parfrey_etal_2015}. At the same time, if the coronal loops encircling the inner one -- i.e., the one that couples the disk to the BH -- are too large compared to the inner loop, they can squeeze the otherwise jet-like outflow, disrupting it through kink instabilities \citep{yuan_etal_2019, mahlmann_etal_2020}.

While the force-free approximation works well to describe the corona, at least outside of reconnecting current sheets, it breaks down inside the heavy, geometrically thin accretion disk, where gravitational and pressure forces become important. 
This is why, in the force-free studies of \citet{parfrey_etal_2015}, \citet{yuan_etal_2019} and \citet{mahlmann_etal_2020}, accretion is prescribed as a simplified boundary condition rather than as a dynamical disk.
The coronal magnetic loops are simply attached to the equator by hand, and inward footpoint dragging is imposed as a proxy for accretion.

To model the disk properly, one must use a more general plasma framework, like magnetohydrodynamics (MHD), that can capture the non-negligible disk-plasma inertia. However, even for MHD simulations, realizing the razor-thin disk limit is numerically challenging since it imposes prohibitive resolution requirements. Nevertheless, recent works have made great strides toward the thin-disk regime. One recent result is that thin disks can efficiently transport magnetic flux inwards in radius \citep{liska_etal_2019, jacquemin-ide_etal_2021, scepi_etal_2024}, which defies preceding theoretical expectations \citep{lubow_etal_1994}. Moreover, even though an efficient disk dynamo has not been observed in the thin-disk limit \citep{musoke_etal_2023}, state-of-the-art simulations of thicker disks \citep{liska_etal_2020} exhibit the generation of large-scale magnetic loops that are accreted inward and fed to the BH \citep{jacquemin-ide_etal_2024}. Even in the thick-disk regime, when such large loops accrete onto the BH, they result in efficient BZ jet activation \citep{chashkina_etal_2021}. Taken together, MHD studies appear to validate, at least in part, the simple prescription of inward magnetic loop dragging adopted by force-free models \citep{parfrey_etal_2015, yuan_etal_2019, mahlmann_etal_2020}. They further suggest that, even if the disk is, in reality, thicker, and the magnetization contrast less pronounced between disk and corona, the basic looped magnetic structure falling onto the BH may still be preserved \citep{chashkina_etal_2021, jacquemin-ide_etal_2024}. 

We will therefore assume, in this work, that the basic magnetic configuration and dynamics of the corona can be studied simply by imposing magnetic footpoint motion as an equatorial boundary condition. 
We will, in addition, attempt to model this problem in a way that retains one major element that neither force-free electrodynamics nor MHD can capture: self-consistent energetics. In particular, we aim for the electromagnetic energy dissipated through reconnection, as well as the reconnection rate itself, to be governed by first-principles plasma physics as opposed, as in MHD or force-free models, to numerical prescriptions. 
Such self-consistency is absolutely necessary to capture reconnection-powered particle acceleration and radiation, and it is also important to ascertain the role of reconnection in regulating the global coronal structure and BZ jet formation. This is because the rate at which reconnection occurs influences how it interacts with the other main effect, (mostly Keplerian) magnetic footpoint motion, to shape the coronal loops \citep[][]{uzdensky_goodman_2008}. 
To achieve our targeted self-consistent energetics, we will perform simulations based on the particle-in-cell (PIC) technique \citep{birdsall_langdon_1991}.

Indeed, because of its attractive self-consistency property, the PIC method has already been used to simulate local models of reconnection in regimes relevant to ADCs \citep{werner_etal_2019, sironi_beloborodov_2020, mehlhaff_etal_2021, mehlhaff_etal_2024, sridhar_etal_2021, sridhar_etal_2023, gupta_etal_2024}. However, we are here keenly interested in the impact of reconnection on the global magnetic configuration, which requires us to perform global simulations. Global PIC models featuring a magnetic coupling between a central spinning BH and its accretion disk have recently been presented by \citet{elmellah_etal_2022} and \citet{elmellah_etal_2023}.
However, their work did not include advection of multiple loops onto the BH.
We aim to fold in this fundamental aspect, and so to unveil how reconnection regulates particle acceleration, coronal structure, and jet formation throughout the loop-accretion cycle.
In this first paper, we focus on the corona-jet interaction in the regime \citep[uncovered by previous force-free works;][]{uzdensky_2005, parfrey_etal_2015, yuan_etal_2019, mahlmann_etal_2020} where the loop size is much larger than the BH, leading to efficient jet launch; in a companion publication \citep{crinquand_etal_inprep}, we tackle the complementary regime where the loops are smaller, and the main reconnection dynamics come from the corona interacting with itself.

This paper is structured as follows. In Section~\ref{sec:setup}, we describe our simulation setup, which is heavily inspired by the aforementioned force-free models \citep{parfrey_etal_2015, yuan_etal_2019, mahlmann_etal_2020}. In Section~\ref{sec:results}, we present the results from our simulation, focusing on those aspects for which the PIC approach is uniquely adapted: the global energetics and (in consequence) the radiative signatures. We discuss implications of our findings for ADCs in the context of XRBs, changing-look active galactic nuclei, Sagittarius A*, and the striped jet model of \citet{giannios_uzdensky_2019} in Section~\ref{sec:discussion}. We summarize our main findings in Section~\ref{sec:conclusions}.

\section{Setup}
\label{sec:setup}
\begin{figure}
    \centering
    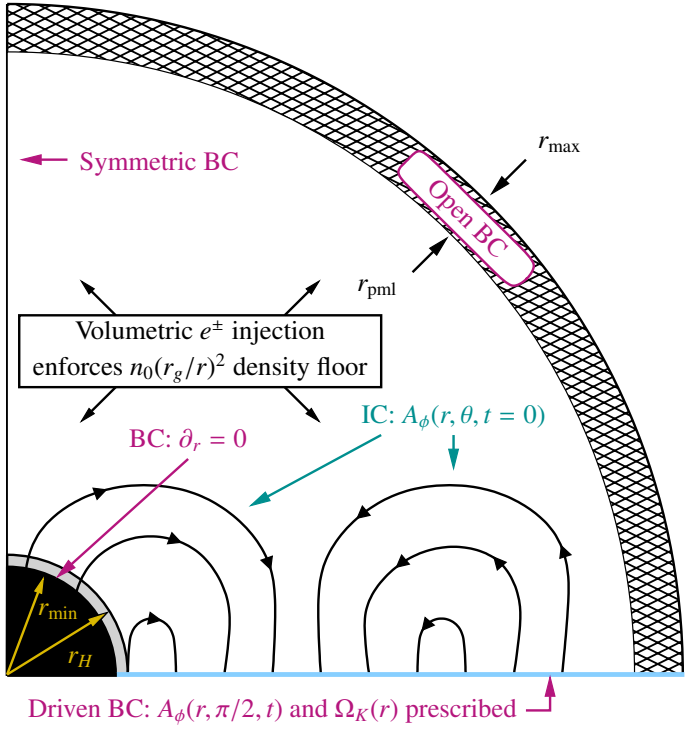
    \caption{Initial conditions (IC) and boundary conditions (BC) of our simulation setup. Sketch is not to scale.}
    \label{fig:setup}
\end{figure}

We present axisymmetric pair-plasma simulations using the general relativistic PIC (GRPIC) code \codename{grzeltron}\citep{parfrey_etal_2019}.
Our simulation setup, sketched in Fig.~\ref{fig:setup}, closely resembles that used by \citet{parfrey_etal_2015} for their force-free modeling.
Here, we summarize the main qualitative features of our setup; we detail technical aspects in the subsections that follow. 

We simulate the upper-half space near a Kerr BH. To model the plasma corona, we attach a tightly packed train of ultramagnetized, purely poloidal (i.e., with no initial azimuthal component) magnetic loops to the simulation equatorial boundary. At the simulation outset, we begin dragging the loop footpoints inwards in radius, while also forcing them into Keplerian rotation in the azimuthal direction. The Keplerian motion is prograde: aligned with the BH spin axis. The rotational shear across loop footpoints builds up an azimuthal component in the magnetic field, magnetically pressurizing the loops and, hence, causing them to inflate towards infinity. Simultaneously, the inward field-line dragging feeds magnetic flux to the BH, enforcing a coupling with the corona and allowing a BZ jet to be launched. 

We choose a rather extreme dimensionless Kerr spin value of~$a=0.99$. While we suspect that our results would hold for a broad range of (prograde) spins, the near-extreme case provides a number of numerical conveniences. For example, it extremizes the rotation of field lines threading the BH horizon and thus renders the closed field-line region smaller. This makes it easier to fit all of the regions of distinct magnetic topology into our simulation box \citep{elmellah_etal_2022}. In addition, a near-extreme spin tightens the radius,~$r_{\rm ISCO}$, of the innermost stable circular orbit. This limits the zone between the event horizon and~$r_{\rm ISCO}$ where our simplified accretion prescription is least realistic.

\subsection{The 3+1 formalism}
Our simulations adopt a~$3+1$ formalism, in which the general relativistic line element is expressed as
\begin{align}
    \dif s^2 &= g_{\mu\nu} \dif x^\mu \dif x^\nu \notag \\
             &= (\beta_l \beta^l - \alpha^2) c^2 \dif t^2 + 2 \beta_i \dif x^i c \dif t + h_{ij} \dif x^i \dif x^j \, ,
    \label{eq:kerr}
\end{align}
Here,~$\alpha$ is the lapse function,~$\beta_i$ are the shift vectors,~$h_{ij} = g_{ij}$ is the spatial part of the metric, Greek indices span space and time (taking values 0-3), Latin indices span only space (values 1-3), and the Einstein summation convention applies. The lapse function and shift vectors define fiducial observers (FIDOs), with four-velocities
\begin{align}
    k^\mu = g^{\mu\nu} k_\nu = g^{\mu\nu} (-\alpha, 0, 0, 0)_\nu = (1, -\beta^1, -\beta^2, -\beta^3)/\alpha \,
    \label{eq:fido4v}
\end{align}
normalized such that~$k^\mu k_\mu = -1$. Unless otherwise stated, we report all simulated quantities as measured by FIDOs.

Throughout this work, we adopt a Kerr spacetime expressed via horizon-penetrating spherical Kerr-Schild coordinates,~$(x^0,x^1,x^2,x^3) = (t,r,\theta,\phi)$. In these coordinates, the metric reads
\begin{align}
    \alpha  &= \frac{1}{\sqrt{1+z}} & \beta^r &= \frac{z}{1+z} & \beta^\theta &= \beta^\phi = 0 \notag \\
    h_{rr}  &= 1+z & h_{\theta\theta} &= \Sigma & h_{\phi\phi} &= \frac{A \sin^2 \theta}{\Sigma} \notag \\
    h_{r\phi} &= -a r_g \sin^2 \theta (1+z) & h_{r\theta} &= h_{\theta\phi} = 0 \, ,
    \label{eq:ksmetric}
\end{align}
where
\begin{align}
    \Sigma &\equiv r^2 + a^2 r_g^2 \cos^2 \theta & z &\equiv 2 r_g r / \Sigma \notag \\
    A &\equiv (r^2 + a^2 r_g^2)^2 - a^2 r_g^2 \Delta \sin^2 \theta & \Delta &\equiv r^2 - 2 r r_g + a^2 r_g^2 \, ,
    \label{eq:ksauxiliary}
\end{align}
$r_g = GM / c^2$ is the gravitational radius, and~$M$ is the BH mass. The determinant of the spatial metric is~$h = \Sigma^2 (1 + z) \sin^2 \theta$.

Maxwell's equations can be stated covariantly as \citep{jackson_1975}
\begin{align}
    \nabla_\mu F^{\nu\mu} &= \frac{4 \pi}{c} I^\nu \quad \mathrm{and} \notag \\
    \nabla_\mu {\ast} F^{\nu\mu} &= 0 \, ,
    \label{eq:maxcov}
\end{align}
where~$\nabla_\mu$ is the covariant derivative,~$I^\mu$ is the four-current density, and~$F^{\mu\nu}$ and~${\ast} F^{\mu\nu}$ are, respectively, the electromagnetic field tensor and its Hodge dual.
In the~$3+1$ formalism adopted in this paper, equations~(\ref{eq:maxcov}) are recast to closely resemble their flat-space form as \citep{komissarov_2004}
\begin{align}
    \partial_t \mathbf{B} &= -c \mathbf{\nabla} \times \vec{E}, \notag \\
    \partial_t \mathbf{D} &= c \mathbf{\nabla} \times \vec{H} - 4 \pi \vec{J}, \notag \\
    \mathbf{\nabla} \cdot \mathbf{D} &= 4 \pi \rho, \quad \mathrm{and} \notag \\
    \mathbf{\nabla} \cdot \mathbf{B} &= 0 \, ,
    \label{eq:maxwell}
\end{align}
where we have assumed a stationary metric.
Here,~$D^i = F^{i\nu} k_\nu = -\alpha F^{i0}$ and~$B^i = -{\ast} F^{i\nu} k_\nu = \alpha {\ast} F^{i0}$ are, respectively, the electric and magnetic fields observed by FIDOs;~$\rho=-I^\mu k_\mu / c$ is the FIDO-observed charge density,~$J^i$ is a current density\footnote{The current density~$J^i$ appearing in equations~(\ref{eq:maxwell}) is not that observed by FIDOs. Rather, the FIDO-observed current density is~$j^i = I^i + (I^\mu k_\mu) k^i = (J^i + c \rho \beta^i) / \alpha$.} given in terms of the timelike Killing vector~$\zeta^\mu = (1,0,0,0)$ of the metric as~$J^i = (I^\mu \zeta^i - I^i \zeta^\mu) k_\mu = \alpha I^i$; and~$\mathbf{E}$ and~$\mathbf{H}$ are the auxiliary fields 
\begin{align}
    \mathbf{E} &= \alpha \mathbf{D} + \mathbf{\beta} \times \mathbf{B} \quad \rm and \notag \\
    \mathbf{H} &= \alpha \mathbf{B} - \mathbf{\beta} \times \mathbf{D} \, .
    \label{eq:heconst}
\end{align}

\subsection{Initial conditions}
\label{sec:ics}
Our simulations start in vacuum with~$D^r = D^\theta = D^\phi = B^\phi = 0$. We set the initial poloidal components of the magnetic field,~$B^r$ and~$B^\theta$, according to a prescribed magnetic flux function~$A_\phi(r,\theta,t=0)$ such that
\begin{align}
    B^r(r,\theta,0) &= \frac{1}{\sqrt{h}} \partial_{\theta} A_\phi(r,\theta,0) \quad \mathrm{and} \notag \\
    B^\theta(r,\theta,0) &= -\frac{1}{\sqrt{h}} \partial_{r} A_\phi(r,\theta,0) \, .
    \label{eq:Bpolinit}
\end{align}
We use this prescription to thread the equatorial boundary with a sequence of magnetic loops, as shown in Fig.~\ref{fig:setup}. 

We choose the functional form of~$A_\phi(r,\theta,0)$ so that the loops all: have approximately the same diameter,~$\lambda = 10r_g$; carry the same amount of magnetic flux; and alternate in magnetic polarity (the direction that the magnetic field circulates around each loop). These properties can all be achieved by selecting~$A_\phi$ on the equator as
\begin{align}
    A_{\phi}(r,\pi/2,0) = \begin{cases}
        B_0 r_g^2 \sin ( \pi r / \lambda ) , & r < 20 r_g \\
        0 ,                                     & \mathrm{otherwise}
    \end{cases} \, ,
    \label{eq:aphiinit_eq}
\end{align}
and extending this into the rest of the simulation domain by requiring the resulting magnetic field to be potential.
As a simplification, instead of using the exact potential-field that results from condition~(\ref{eq:aphiinit_eq}), we adopt the solution that would follow if~$r$,~$\theta,$ and~$\phi$ were the standard flat-space spherical coordinates:
\begin{align}
    A_\phi(r,\theta,0) = r \sin \theta \int_0^\infty B(\kappa) J_{1} (\kappa r \sin \theta) e^{-\kappa r \cos \theta} \dif \kappa ,
    \label{eq:aphiinit}
\end{align}
where
\begin{align}
    B(\kappa) = \kappa \int_0^\infty A_\phi (R,\pi/2,0) J_{1}(\kappa R) \dif R \,
    \label{eq:bkappadef}
\end{align}
and~$J_{1}$ is a Bessel function of the first kind. Because expression~(\ref{eq:aphiinit}) ignores spacetime curvature, our initial magnetic field is not formally current free. This results in a mild transient that quickly evacuates our simulation in the first several~$r_g / c$ -- well in advance of the times we analyze in this work. Our main objective in using expression~(\ref{eq:aphiinit}) is not to start from a formal equilibrium, but rather to embed our simulation with an initial looped magnetic topology (Fig.~\ref{fig:setup}). 

\subsection{Boundary conditions}
\label{sec:bcs}
At the~$\theta=0$ and~$r=r_{\rm max}$ boundaries, we employ axisymmetric and open boundary conditions, respectively. For the open boundary at~$r=r_{\rm max}$, this involves setting up a perfectly matched layer \citep[PML;][]{birdsall_langdon_1991, berenger_1994, cerutti_etal_2015} where the fields are smoothly damped to zero over a finite shell extending from~$r=r_{\rm pml}=27r_g$ to~$r=r_{\rm max}=30r_g$. Particles that reach~$r_{\rm max}$ are deleted.

We place the inner radius,~$r=r_{\rm min}$, of our simulation underneath the BH event horizon,~$\rh = r_g (1 + \sqrt{1 - a^2}) \simeq 1.14 r_g$, setting~$r_{\rm min} = 0.9 \rh$. This conveniently renders the numerics insensitive to how we handle the~$r=r_{\rm min}$ surface. For simplicity, we enforce a zero radial derivative here on~$\mathbf{D}$,~$\mathbf{E}$,~$\mathbf{B}$, and~$\mathbf{H}$, and we delete all particles that touch~$r_{\rm min}$.

The main nontrivial boundary condition in our simulations is applied across a thin, one-cell thick disk lining the equator. This is the main control surface that we use to drive the simulation dynamics.
Beginning at time~$t = 0$, we throw the footpoints of magnetic field lines threading the disk into rotation at the local Keplerian angular velocity. We do this by applying a poloidal electric field,~$E_r = \sqrt{h} \Omega_{\rm K} B^\theta$ and~$E_\theta = -\sqrt{h} \Omega_{\rm K} B^r$, where
\begin{align}
    \Omega_{\rm K} = \sqrt{\frac{G M}{r_g^3}} \frac{1}{(r \sin \theta / r_g)^{3/2} + a} \, ,
    \label{eq:omegak}
\end{align}
and~$r \sin \theta \simeq r$ is the distance to the symmetry axis.

Besides Keplerian rotation, we also prescribe a time-varying~$A_\phi$ inside the equatorial disk, continuous with our initial condition~(\ref{eq:aphiinit_eq}), of the form
\begin{align}
    A_{\phi}(r,\pi/2,t) = \begin{cases}
        B_0 r_g^2 \sin \left( \frac{\pi}{\lambda} (r+v_0 t) \right) , & r < 20 r_g \\
        0 ,                                     & \mathrm{otherwise}
    \end{cases}
    \label{eq:aphibc} \, .
\end{align}
This pulls the footpoints of the magnetic field lines threading the equator inward at speed~$v_0 = 0.01 c$, mimicking the dragging of magnetic loops by an underlying accretion disk.
Our intent in choosing~$v_0 = 0.01 c$ is not to model a real accretion speed. Instead, we merely aim to have the accretion time for a single loop,~$t_{\rm acc} = \lambda / v_0$, much longer than all the timescales governing the coronal dynamics, the longest of which is the time it takes to reconnect the flux in one loop,~$t_{\rm rec} \sim \lambda / \beta_{\rm rec} c$, where~$\beta_{\rm rec} \simeq 0.1$ is a typical relativistic magnetic reconnection rate. As long as the hierarchy~$t_{\rm acc} \gg t_{\rm rec}$ (i.e.,~$v_0 \ll \beta_{\rm rec} c \simeq 0.1 c$) is respected, the imposed inward field-line dragging just causes the simulation to pass through a sequence of quasisteady states, each one corresponding to a different set of magnetic footpoint locations.

\subsection{Plasma supply}
\label{sec:supply}
For simplicity, we adopt an \textit{ad hoc} volumetric plasma injection scheme. At every timestep, for any cell in which the overall (electron+positron) plasma number density,~$n$, falls below a prescribed floor of
\begin{align}
    n_{\rm fl} = n_0 \left( \frac{r_g}{r} \right)^2 \, ,
    \label{eq:nfloor}
\end{align}
we inject electron-positron pairs to bring the density back up to~$n_{\rm fl}$.
Plasma is injected with zero mean FIDO-measured velocity and with a FIDO-frame temperature~$kT=m_e c^2$. The power-law~$n_{\rm fl} \propto r^{-2}$ maintains a roughly constant (modulo sinusoidal modulations) equatorial plasma magnetization,~$\sigma \equiv B_i B^i / 4 \pi n m_e c^2$, since, from equations~(\ref{eq:Bpolinit}) and~(\ref{eq:aphibc}), the square of the vertical field threading the disk,~$\sim h_{\theta\theta} B^\theta B^\theta$, also decays with a~$1/r^2$ profile. We weight the injected particles to achieve a simulation average of about 20 macroparticles per grid cell.

In addition, we apply synchrotron radiative cooling to the plasma. This ensures that the high-energy radiation that we diagnose in Section~\ref{sec:synthobs} is produced by particles that are rapidly cooled, which we expect in highly magnetized BH ADCs \citep{beloborodov_2017}. We relegate a study involving inverse Compton cooling, which is also potentially important in this context, to a future work. We calibrate synchrotron cooling so that all particles with relativistic (FIDO-measured) Lorentz factors~$\gamma \gg 1$ cool down to transrelativistic energies,~$\gamma \gtrsim 1$, within a characteristic lightcrossing time,~$r_g/c$, of the BH.

\subsection{Parameter values and grid selection}
\label{sec:params}
We choose the parameters~$B_0$ and~$n_0$ in conjunction with our simulation grid so that the plasma in our simulation (at least outside of reconnection sites) is highly magnetized and force-free, as expected in BH ADCs. In addition, in order to avoid spurious numerical effects, we need to ensure that important plasma microscales are resolved by our simulation grid. These objectives requires us to meet three simultaneous conditions:
\begin{enumerate}
    \item A healthy plasma supply, with enough particles to carry the current necessary for the force-free electromagnetic fields.
     \label{en:condsupply}
    \item A high magnetization~$\sigma \gg 1$ (except potentially inside reconnecting current sheets).
     \label{en:condmag}
    \item A grid that is fine enough to resolve the plasma skin depth everywhere.
      \label{en:condres}
\end{enumerate}
Conditions~\ref{en:condsupply} and~\ref{en:condres} are most stringent near the BH horizon, where the plasma is densest, the magnetic field strongest, and the rotation most rapid (i.e., shortest skin depth and highest demand for electrical charge). Therefore, meeting these two conditions closest to the BH suffices to meet them throughout our simulation. Condition~\ref{en:condmag}, on the other hand, tends to become more strained at large~$r$ as~$\sigma$ decays. To ensure~$\sigma \gg 1$ at all radii, we set the fiducial magnetization near the BH horizon,~$\sigma_0 = B_0^2 / 4 \pi n_0 m_e c^2$, as high as we can afford while respecting conditions~\ref{en:condsupply} and~\ref{en:condres}. We then check \textit{a posteriori} that~$\sigma \gg 1$ throughout our simulation box.

To ensure adequate plasma supply, we choose~$n_0$ as a multiple,~$\eta$, of the fiducial Goldreich-Julian density,~$\ngjfid = B_0 \ombh / 4 \pi c e$, where~$\ombh = a c / 2 \rh$ is the BH angular velocity. We set~$\eta \equiv n_0 / \ngjfid = 3$. This takes care of condition~\ref{en:condsupply}.

To handle conditions~\ref{en:condmag} and~\ref{en:condres}, we recast the magnetic field in terms of the nominal gyroradius~$\rho_0 \equiv m_e c^2 / e B_0$, which permits us to rewrite the horizon-scale skin depth,~$\dfid = (m_e c^2 / 4 \pi n_0 e^2)^{1/2}$, and magnetization,~$\sigma_0 = B_0^2 / 4 \pi n_0 m_e c^2$, as
\begin{align}
    \dfid^2 = \frac{1}{\multiplicity} \frac{m_e c^2}{e B_0} \frac{c}{\ombh} = \frac{2 \rho_0 r_g}{\multiplicity} \left( \frac{1 + \sqrt{1 - a^2}}{a} \right) \simeq 0.8 \rho_0 r_g
    \label{eq:dfidlong}
\end{align}
and
\begin{align}
    \sigma_0 = \frac{1}{\multiplicity} \frac{e B_0}{m_e c^2} \frac{c}{\ombh} = \frac{2}{\multiplicity} \frac{r_g}{\rho_0} \left( \frac{1 + \sqrt{1-a^2}}{a} \right) \simeq 0.8 \frac{r_g}{\rho_0} .
    \label{eq:sigfidlong}
\end{align}
The values~$a=0.99$ and~$\multiplicity=3$ are assumed at the ends of both lines. 
These expressions show that the need for a well-resolved skin depth is in direct tension with that for a high magnetization, linking high~$\sigma$ directly to computational cost.
To ease the numerical burden, we employ a logarithmically stretched grid, keeping~$\Delta (\ln r)$ constant, which concentrates resolution toward the BH. We then set~$\rho_0 = r_g/4000$, yielding~$\dfid \simeq 0.01 r_g$ and~$\sigma_0 \simeq 3000$ -- sufficient to maintain~$\sigma \gg 1$ throughout the simulation domain -- and we lay out~$N_r = 1024$ cells in~$r$ and~$N_\theta = 512$ cells in~$\theta$ (with~$\Delta \theta$ constant). This grid results in a skin-depth resolution of~$d_{e0}/\Delta r \simeq 4$ at~$r = \rh$.

\section{Results}
\label{sec:results}
\begin{figure*}
    \centering
    \includegraphics{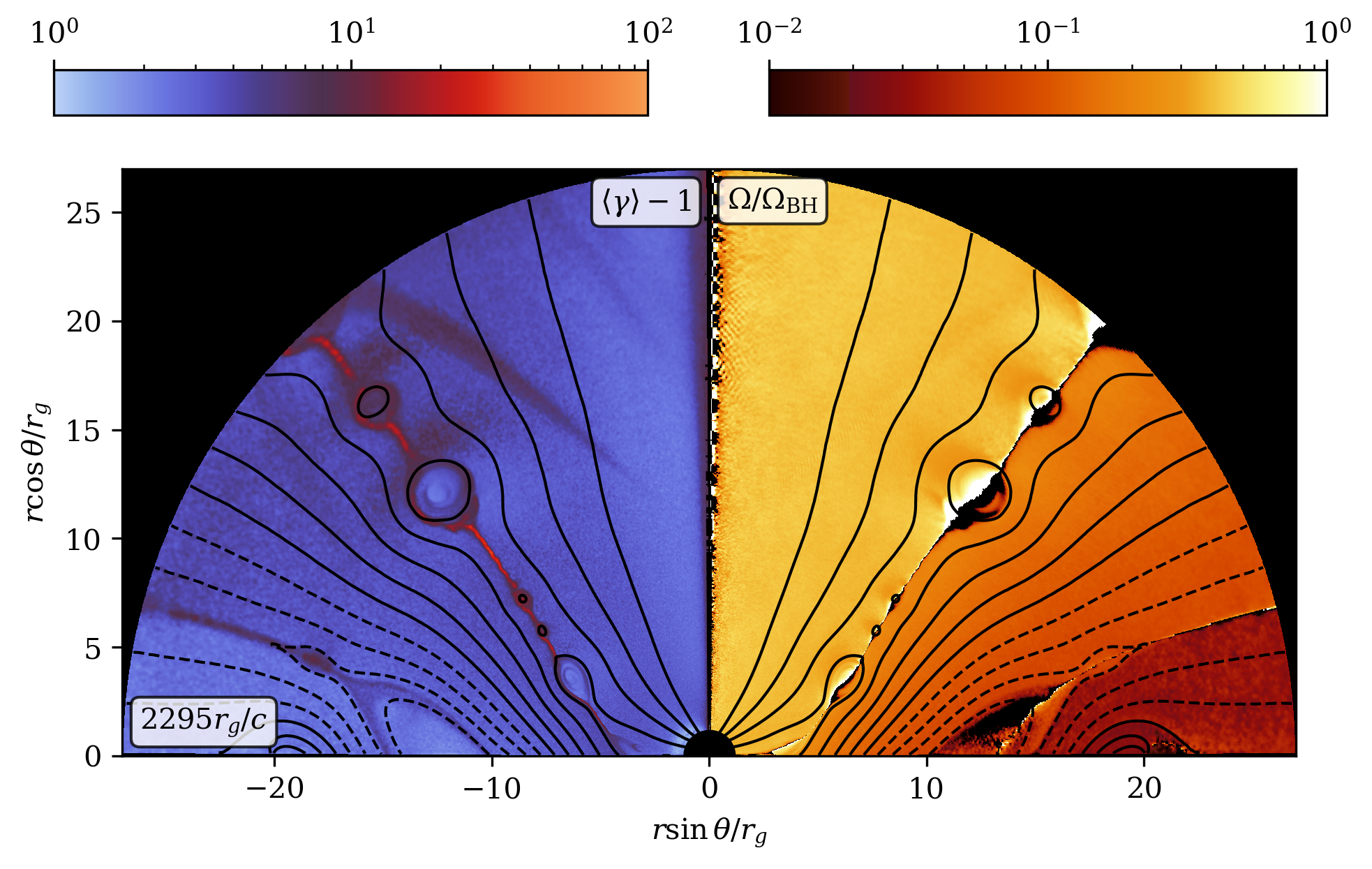}
    \caption{Snapshot of the simulation analyzed in this work. Left: Local average particle Lorentz factor~$\langle \gamma \rangle$. Right: Angular velocity,~$\Omega = -E_\theta / \sqrt{h} B^r$, of magnetic field lines. Rotational shear along magnetic field lines causes the initially closed magnetic topology to partially open. A fully open topology is prevented by reconnection, which allows some closed loops to persist. Reconnecting current sheets are visible as sights of particle acceleration (left) and discontinuities in field-line angular velocity (right). A movie of this simulation is available as at: \url{https://youtu.be/G1q14VKcmQM?si=lBj6mDrctv4lChhh}.}
    \label{fig:lfomega}
\end{figure*}

\subsection{Overall dynamics}
\label{sec:overall}
A snapshot showcasing the main dynamical aspects of our simulation is presented in Fig.~\ref{fig:lfomega}. Rotational shear imposed by the equatorial disk and BH builds up an azimuthal component in the magnetic field of closed loops, pressurizing and inflating them towards infinity. As the loops inflate, sharp, current-sheet discontinuities develop in the magnetic field and begin reconnecting. Reconnecting current sheets separate differentially rotating open field lines, thus appearing as discontinuities in the map of field-line angular velocity,~$\Omega = -E_\theta / \sqrt{h} B^r$, in Fig.~\ref{fig:lfomega}.
 
The competition between rotational shear and reconnection sets the size of the open field-line regions \citep{uzdensky_goodman_2008}. On the one hand, rotation inflates magnetic loops, pumping energy into the field and promoting a progressively higher-energy open configuration \citep[with the theoretical maximum-energy limit being a fully open field;][]{aly_1984, aly_1991, sturrock_1991}. On the other hand, reconnection allows the field to snap back, permitting closed field lines to persist and preventing a fully open (maximum-energy) configuration from being reached. An advantage of our fully kinetic model is that the reconnection rate is set self-consistently, and, thus, so too are the relative sizes of the open and closed field-line regions. Our finite simulation domain can potentially artificially accentuate open field-line regions, however, by cutting off reconnection at the outer boundary.

Despite rapid reconnection, prominent bundles of open magnetic flux persist, transmitting Poynting flux from their footpoints toward infinity, either in the form of a BZ jet funnel (for field lines attached to the BH; \citealt{blandford_znajek_1977}) or in the form of a force-free wind (for field lines attached to the disk; \citealt{blandford_1976}). This transmission is imperfect, however, because some of the injected Poynting flux gets consumed by reconnection. The dissipated energy is used to accelerate particles and, thus, current sheets appear as thin, red-hot regions in Fig.~\ref{fig:lfomega}. The energy budget of this system, especially the efficiency with which Poynting flux is transmitted to infinity versus consumed by reconnection, is the subject of Section~\ref{sec:enbudget}. The observable synchrotron radiation produced by the accelerated particles is the focus of section~\ref{sec:synthobs}.

The open flux bundles in Fig.~\ref{fig:lfomega} have a nearly radial, monopolar shape at large~$r$.
This is similar to the final field structure predicted by earlier studies of axisymmetric force-free equilibria \citep{barnes_sturrock_1972, lynden-bell_boily_1994}, wherein an initially closed magnetic field, line-tied to a differentially rotating disk, is quasistatically twisted open. Far away from the disk region where the flux is tied, the open field becomes nearly radial. Modulo reversals across current sheets, the strength of this field is approximately that of a monopole with flux equal to the total unsigned open flux threading the disk.

The sites of most intense particle acceleration in Fig.~\ref{fig:lfomega} coincide with reconnection along field lines coupling the disk to the BH. This indicates the BH-disk interaction as an important driver of the dynamical activity. Reconnection still occurs on disk-disk field lines, but it is less intense, owing to the lower field strength and rotational shear.

Over long timescales, visible in the animated movie to accompany Fig.~\ref{fig:lfomega}, one can see the slow advection, at speed~$v_0 = 0.01 c$, of the magnetic footpoints toward the BH. Once the footpoints of the innermost loop get pushed onto the horizon, vigorous reconnection begins to annihilate the loop, ejecting its flux away in the form of a rapid-fire barrage of plasmoids launched towards infinity. 

Figure~\ref{fig:lfomega} and its accompanying film also show, at the outer disk radii (close to~$r = 20 r_g$) the formation of new magnetic loops. This is a self-consistent byproduct of our dynamical driving (discussed in section~\ref{sec:bcs}) of~$A_{\phi}$ on the equatorial boundary of our simulation. Because new loops are formed \textit{in situ}, we are able to simulate as many loop advection and ejection cycles as we wish without needing to fit all of the loops onto the equatorial disk initially. We are therefore able to present three such cycles, though we thread the disk with only two loops at~$t=0$.

\subsection{Energy budget}
\label{sec:enbudget}
Here, we analyze the energy budget in our system. First, in Section~\ref{sec:enbudgetmeasure}, we empirically track the fate of the Poynting flux injected by the equatorial disk and BH. We quantify how much of this Poynting flux is transmitted through our simulation versus converted to particle kinetic energy, and we characterize the efficiency with which a BZ jet is launched. Subsequently, in Section~\ref{sec:enbudgetmodel}, we present a toy phenomenological model to explain the simulated dissipation in terms of the rate,~$\beta_{\rm rec}$, of relativistic magnetic reconnection. We summarize the key takeaways from Sections~\ref{sec:enbudgetmeasure} and~\ref{sec:enbudgetmodel} in Section~\ref{sec:enbudgetsummarize}.

\subsubsection{Measurements from simulation}
\label{sec:enbudgetmeasure}
\begin{figure}
    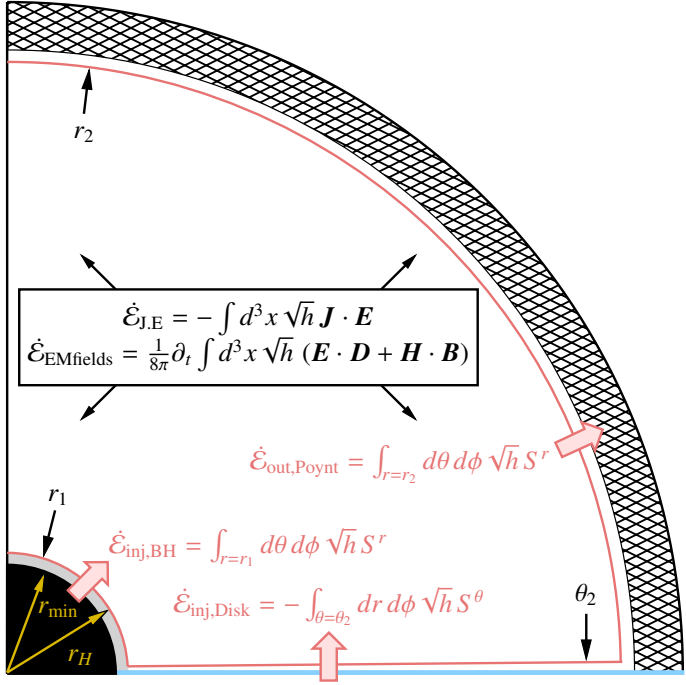
    \caption{The integration surface (pink) over which we evaluate Poynting's Theorem extends from~$r_1 = \rh$ up to~$r_2 = 25 r_g < r_{\rm pml}$ and from~$\theta=0$ to~$\theta_2 = 89^\circ$ (just outside the equatorial disk). We decompose contributions to Poynting's Theorem into the terms shown. In the expressions,~$\vec{S} \equiv c \vec{E} \times \vec{H} / 4 \pi$. We define~$\dot{\mathcal{E}}_{\rm inj,Poynt}\equiv\dot{\mathcal{E}}_{\rm inj,BH}+\dot{\mathcal{E}}_{\rm inj,Disk}$.}
    \label{fig:poyntsurface}
\end{figure}

Our main analysis technique to quantify the energy budget in our system is to evaluate Poynting's Theorem, which can be phrased in the~$3+1$ formalism as \citep{komissarov_2004}
\begin{align}
    \partial_t \left[ \frac{1}{8\pi} \left( \vec{E} \cdot \vec{D} + \vec{H} \cdot \vec{B} \right) \right] + \vec{\nabla} \cdot \left[ \frac{c}{4 \pi} \left( \vec{E} \times \vec{H} \right) \right] = - \vec{J} \cdot \vec{E} \, .
    \label{eq:poynt}
\end{align}
We integrate~(\ref{eq:poynt}) over the surface shown in Fig.~\ref{fig:poyntsurface}, decomposing into contributions from the terms defined in the same figure. 
These terms are: the Poynting flux injected from the disk and BH,~$\dot{\mathcal{E}}_{\rm inj,Poynt}$; the Poynting flux escaping the simulation,~$\dot{\mathcal{E}}_{\rm out,Poynt}$; the Poynting flux consumed to energize the plasma,~$\dot{\mathcal{E}}_{\rm J.E}$; the rate of change of electromagnetic energy,~$\dot{\mathcal{E}}_{\rm EMfields}$; and the Poynting-theorem residual (which should equal zero).
We note that here, as throughout the remainder of the text, we refer to the quantities in Poynting's theorem according to their familiar flat-space names, such as ``Poynting flux'' or ``electromagnetic energy density'', even though equation~(\ref{eq:poynt}) is written in terms of Poynting flux and electromagnetic energy density measured at infinity (as opposed to locally by FIDOs).

\begin{figure*}
    \centering
    \includegraphics{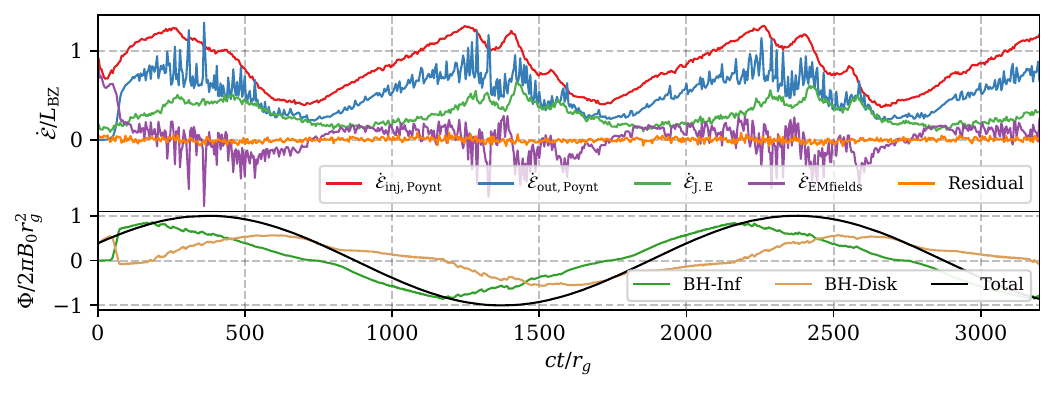}
    \caption{Top: Time series of the terms, defined in Fig.~\ref{fig:poyntsurface}, contributing to Poynting's Theorem. Bottom: Total magnetic flux~(\ref{eq:fluxdef}) piercing the BH horizon, including decomposed contributions~(\ref{eq:fluxdefi}) from open (BH-Inf) and disk-connected (BH-Disk) field lines.}
    \label{fig:enbudget}
\end{figure*}

We present time series of the energy fluxes defined above in Fig.~\ref{fig:enbudget}. We also present the time-averaged values of these fluxes in Table~\ref{table:edotavg}.
\begin{table}
    \centering
    \begin{threeparttable}
        \caption{Average values of the time series in Fig.~\ref{fig:enbudget}. Averages are taken over~$200 \leq c t / r_g \leq 3200$, which spans three loop advection periods and ignores the initial transient. The residual is zero to the reported precision; the apparent error in the first column is due only to rounding.}
        \label{table:edotavg}
        \begin{tabularx}{0.9\columnwidth}{p{0.28\linewidth} p{0.28\linewidth} p{0.28\linewidth}}
            \toprule
            & $\langle \dot{\mathcal{E}} \rangle / L_{\rm BZ}$ & $\langle \dot{\mathcal{E}} \rangle / \langle \dot{\mathcal{E}}_{\rm inj,Poynt} \rangle$ \\
            \midrule
            $\langle \dot{\mathcal{E}}_{\rm inj,Poynt} \rangle$ & 0.82 & 1.00 \\
            $\langle \dot{\mathcal{E}}_{\rm out,Poynt} \rangle$ & 0.53 & 0.64 \\
            $\langle \dot{\mathcal{E}}_{\rm E.J} \rangle$ & 0.30 & 0.36 \\
            $\langle \dot{\mathcal{E}}_{\rm EMfields} \rangle$ & 0.00 & 0.00 \\
            \midrule
            Residual & 0.00 & 0.00 \\
            \bottomrule
        \end{tabularx}
    \end{threeparttable}
\end{table}
In both the figure and table, we define~$L_{\rm BZ} \equiv r_g^4 B_0^2 \Omega_{\rm BH}^2 / 12 c$ as the BZ power extracted from the upper hemisphere of a BH threaded by a monopolar magnetic field of horizon strength~$B_0$ \citep[][]{blandford_znajek_1977, tchekhovskoy_etal_2011, crinquand_etal_2020}. In our simulations, a monopole roughly describes the field lines at large~$r$ as well as the open field lines crossing the horizon.

In Fig.~\ref{fig:enbudget}, the input and output Poynting fluxes evolve roughly in phase, with approximately two thirds of the injected Poynting flux being transmitted out of the simulation box. This agrees with the time-average result from Table~\ref{table:edotavg} that~$\langle \dot{\mathcal{E}}_{\rm out,Poynt} \rangle \simeq 2 \langle \dot{\mathcal{E}}_{\rm inj,Poynt} \rangle / 3$. The remaining one third of the injected electromagnetic energy is used to energize the plasma through magnetic reconnection, with~$\langle \dot{\mathcal{E}}_{\rm J.E} \rangle \simeq \langle \dot{\mathcal{E}}_{\rm inj,Poynt} \rangle /3$.

The contribution from~$\dot{\mathcal{E}}_{\rm EMfields}$ to Poynting's Theorem goes to zero on time average, since we average, by design, over an integer number of advection cycles.
Theoretically, we expect~$\dot{\mathcal{E}}_{\rm EMfields}$ to approach zero at all times, not just on the time average, in the~$v_0 \to 0$ limit. Otherwise, the simulation would not attain a quasisteady state for each magnetic footpoint configuration. We have checked this explicitly by running simulations (not presented) with varying~$v_0$, which show that a smaller value results in a smaller contribution from~$\dot{\mathcal{E}}_{\rm EMfields}$ at any given time to the balance of Poynting's Theorem. The same exercise, on the other hand, does not change the time averages of the various~$\dot{\mathcal{E}}$ terms presented in Table~\ref{table:edotavg}. We therefore conclude that the time-averaged values presented in Table~\ref{table:edotavg} represent the true slow-advection limit.

The bottom panel of Fig.~\ref{fig:enbudget} shows the evolution of the instantaneous magnetic flux,
\begin{align}
    \Phi = 2 \pi \int_0^{\pi/2} B^r \sqrt{h} \dif \theta \, , 
    \label{eq:fluxdef}
\end{align}
piercing the BH upper hemisphere. The~$\Phi$ time series follows a clean sinusoid, reflecting the driving boundary condition~(\ref{eq:aphibc}). The energy budget is highly sensitive to~$\Phi$: more horizon flux generally corresponds to higher energy injection~($\dot{\mathcal{E}}_{\rm inj,Poynt}$), dissipation~($\dot{\mathcal{E}}_{\rm J.E}$), and transmission~($\dot{\mathcal{E}}_{\rm out,Poynt}$). This emphasizes the result already observed in the discussion of Fig.~\ref{fig:lfomega} (Section~\ref{sec:overall}) that the dynamics are strongly driven by the magnetic coupling between the disk and the BH. Maxima in~$|\Phi|$ represent moments where this coupling is quite efficient, powering the most vigorous reconnection and producing coincident maxima in~$\dot{\mathcal{E}}_{\rm J.E}$.

Notably, however, the BH flux,~$\Phi$, is not exactly in phase with the total injected power,~$\dot{\mathcal{E}}_{\rm inj,Poynt}$. Moreover, the latter occasionally exceeds~$L_{\rm BZ}$. This indicates that the energetics of this system are not purely due to the BZ process; an appreciable amount of the injected energy also comes from the disk. To isolate the individual contributions from the BH and disk requires a more detailed energy budget assessment. 

\begin{figure}
    \includegraphics{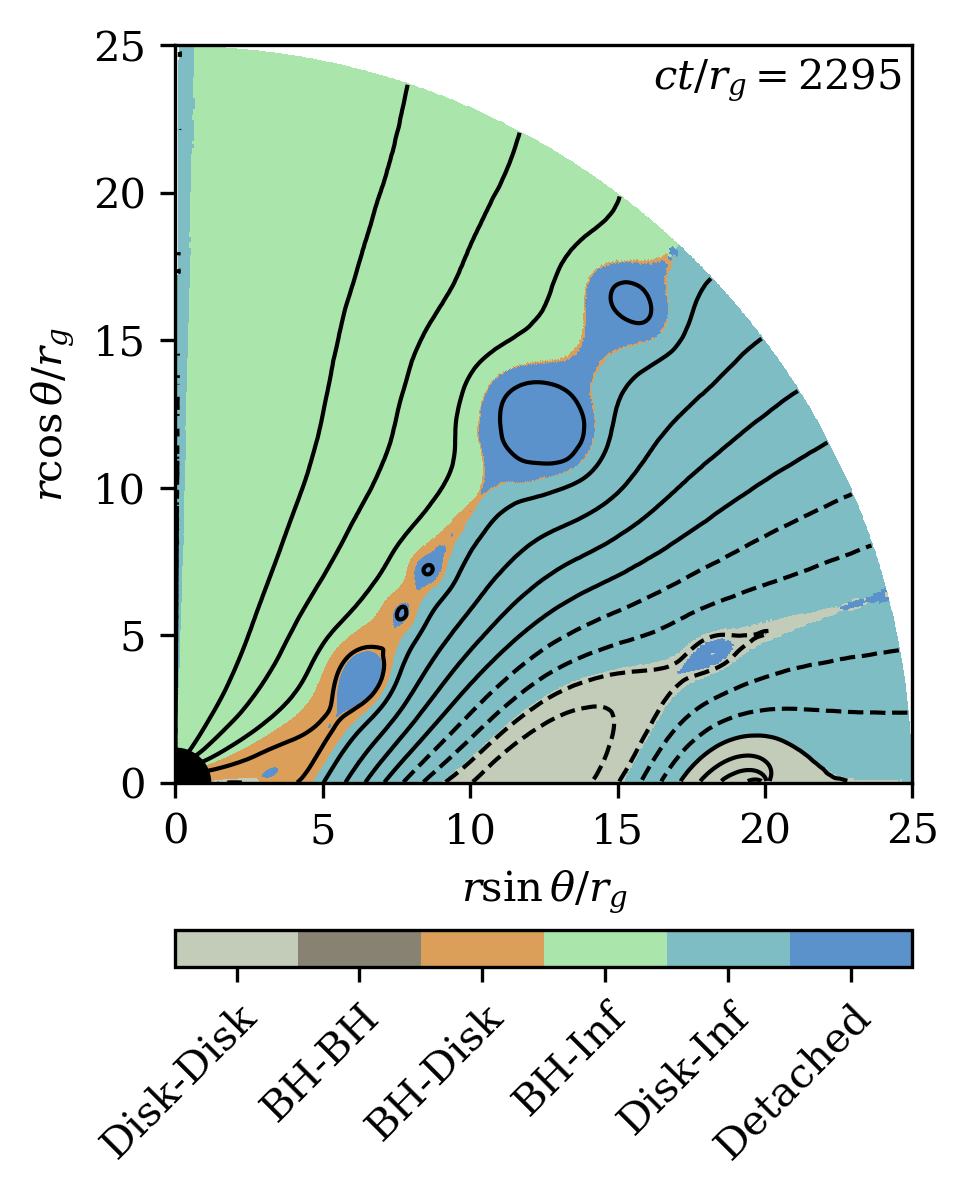}
    \caption{Classification of spatial regions in our simulation domain based on magnetic connectivity. A movie of this simulation is available at: \url{https://youtube.com/shorts/J0gzuw36Qfw?si=qH0-KuSLfMK5RvFI}.}
    \label{fig:maglinks}
\end{figure}

We therefore complement the Poynting Theorem analysis presented above with a more fine-grained view. Here, we decompose the Poynting flux flowing into and out of our integration surface of Fig.~\ref{fig:poyntsurface} into contributions from different field-line bundles. To enable this analysis, we first classify all the points in our simulation domain based on their magnetic connectivity. Points lying on field lines connecting the disk to the BH are labeled ``BH-Disk''; those on field lines coupling the disk to itself are labeled ``Disk-Disk''; points on black-hole ingrown field lines as ``BH-BH''; those inside open magnetic flux tubes anchored to the disk or BH as ``Disk-Inf'' or ``BH-Inf'', respectively; and those on field lines tethered neither to the BH nor to the disk as ``Detached''. This labeling scheme is illustrated in Fig.~\ref{fig:maglinks}.
We note that ingrown BH-BH field lines are highly transient structures in our simulation. They only briefly occur at the moment when the final shred of magnetic flux in the innermost loop is ejected, by reconnection, to infinity: the corresponding morsel of reconnected flux still attached to the BH (the ingrown magnetic hair) quickly sinks through the horizon. In this way, the BH serves as a magnetic flux sink -- a role it plays in addition to injecting electromagnetic energy via the BZ mechanism.

Our magnetic connectivity classification allows us to decompose the magnetic flux threading the BH into its contributions from BH-Inf (open) and BH-Disk (disk-connected) field lines. Formally, we define
\begin{align}
    \Phi_{\bundleidx} = 2 \pi \int_0^{\pi/2} \Theta_{\bundleidx}(\rh,\theta) B^r \sqrt{h} \dif \theta \, , 
    \label{eq:fluxdefi}
\end{align}
where~$\Theta_{\bundleidx}(r,\theta)$ is equal to one if the point~$(r,\theta)$ lies on flux bundle~$\bundleidx$ (e.g.,~$\bundleidx=$ BH-Inf) and zero otherwise. The total horizon flux can be written~$\Phi = \sum_{\bundleidx} \Phi_{\bundleidx} = \Phi_{\rm BH-Inf} + \Phi_{\rm BH-Disk}$. Here, the sum over field-line bundles recovers the only two sets of field lines, BH-Inf and BH-Disk, that, by definition, yield nonzero contributions to~$\Phi$. The lower panel of Fig.~\ref{fig:enbudget} shows time series of~$\Phi_{\rm BH-Inf}$ and~$\Phi_{\rm BH-Disk}$ alongside that of~$\Phi$. Loud periods in the overall dissipation,~$\dot{\mathcal{E}}_{\rm J.E}$, follow peaks in~$|\Phi_{\rm BH-Inf}|$. This shows that the most intense magnetic reconnection in our system preys on the jet funnel, eating away at the BZ jet to power particle acceleration and (as we will see in Section~\ref{sec:synthobs}) bright high-energy emission.

\begin{figure*}
    \centering
    \includegraphics{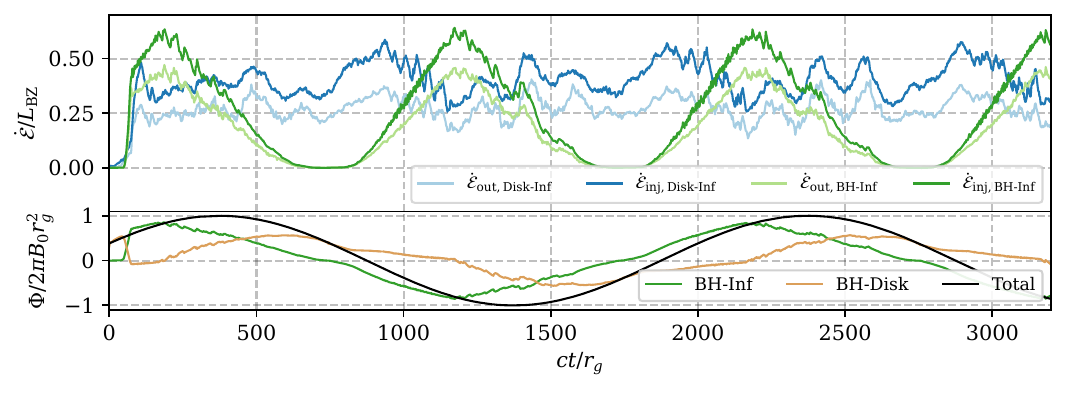}
    \caption{Top: Time series of Poynting flux entering (escaping) our integration surface -- depicted in Fig.~\ref{fig:poyntsurface} -- on open field lines threading the BH,~$\dot{\mathcal{E}}_{\rm inj,BH-Inf}$~($\dot{\mathcal{E}}_{\rm out,BH-Inf}$), and disk,~$\dot{\mathcal{E}}_{\rm inj,Disk-Inf}$~($\dot{\mathcal{E}}_{\rm out,Disk-Inf}$). Bottom: Time series of open (BH-Inf), closed (BH-Disk), and total horizon-piercing magnetic flux as in Fig.~\ref{fig:enbudget}.}
    \label{fig:einoutbhdisk}
\end{figure*}

In Fig.~\ref{fig:einoutbhdisk}, we use our magnetic connectivity classification to analyze the energy flux flowing into and out of our Poynting integration surface (Fig.~\ref{fig:poyntsurface}) on the Disk-Inf and BH-Inf field-line bundles. From here onward, we refer to these two bundles, respectively, as the \textit{disk wind} and the \textit{jet}. Although we focus on just the jet and disk wind here, for completeness, we present in Appendix~\ref{sec:enbudget_full} the energy budget including contributions from all the magnetic connectivity regions.
\begin{table}
    \centering
    \begin{threeparttable}
        \caption{Top: Average values of the time series from Fig.~\ref{fig:einoutbhdisk}. Like Table~\ref{table:edotavg}, averages are taken over~$200 \leq c t / r_g \leq 3200$. Bottom: Efficiency of Poynting flux transmission on BH- and disk-attached flux bundles.}
        \label{table:edotavg_open}
        \begin{tabularx}{0.9\columnwidth}{p{0.28\linewidth} p{0.28\linewidth} p{0.28\linewidth}}
            \toprule
            & $\langle \dot{\mathcal{E}} \rangle / L_{\rm BZ}$ & $\langle \dot{\mathcal{E}} \rangle / \langle \dot{\mathcal{E}}_{\rm inj,Poynt} \rangle$ \\
            \midrule
            $\langle \dot{\mathcal{E}}_{\rm inj,BH-Inf} \rangle$ & 0.25 & 0.30 \\
            $\langle \dot{\mathcal{E}}_{\rm out,BH-Inf} \rangle$ & 0.18 & 0.21 \\
            $\langle \dot{\mathcal{E}}_{\rm inj,Disk-Inf} \rangle$ & 0.41 & 0.49 \\
            $\langle \dot{\mathcal{E}}_{\rm out,Disk-Inf} \rangle$ & 0.27 & 0.33 \\
            \midrule
        \end{tabularx}
        \begin{tabularx}{0.9\columnwidth}{p{0.55\columnwidth} p{0.55\columnwidth}}
            \multicolumn{2}{c}{\quad\quad\quad\quad\quad\quad Transmission efficiencies} \\
            \midrule
            $\langle \dot{\mathcal{E}}_{\rm out,BH-Inf} \rangle / \langle \dot{\mathcal{E}}_{\rm inj,BH-Inf} \rangle $ & \multicolumn{1}{c}{0.71} \\
            $\langle \dot{\mathcal{E}}_{\rm out,Disk-Inf} \rangle / \langle \dot{\mathcal{E}}_{\rm inj,Disk-Inf} \rangle$ & \multicolumn{1}{c}{0.67} \\
            \bottomrule
        \end{tabularx}
    \end{threeparttable}
\end{table}

Figure~\ref{fig:einoutbhdisk} shows that the transmission of Poynting flux through the system, in both the jet and wind, is extremely well correlated with injection from the corresponding energy source: either the BH or the disk. The transmission efficiency for each source is nearly time-independent, with the jet and wind field lines both relaying about two thirds of their initial Poynting flux to the outer edge of the domain. This is, of course, also true of the time-averaged transmission efficiencies reported in Table~\ref{table:edotavg_open}. We further note from Fig.~\ref{fig:einoutbhdisk} that, while the disk wind is roughly stationary, the jet turns on and off with a duty cycle of roughly one half. This permits the jet to temporarily become more powerful than the wind even though the wind carries more Poynting flux on average. 

\begin{figure}
    \centering
    \includegraphics[width=\columnwidth]{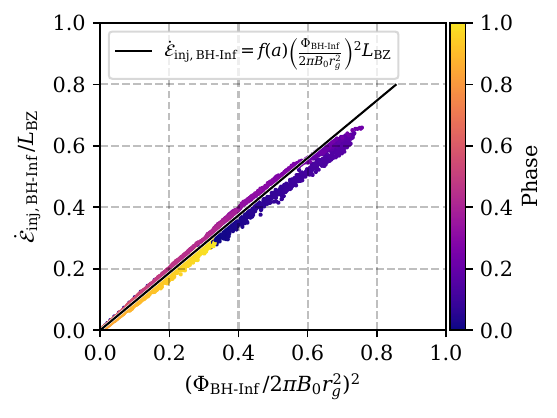}
    \caption{Poynting power injected on, versus square of the magnetic flux comprised by, open field lines threading the BH horizon. Each data point represents one snapshot of the simulation. Color indicates time phase-folded on the loop advection and ejection period,~$1000 r_g/c$. The expected correlation, explained in the text, is drawn in black with~$f(a)$ evaluated as~$f(0.99)=0.93$.}
    \label{fig:bzdemo}
\end{figure}

The lower panel of Fig.~\ref{fig:einoutbhdisk} shows, like Fig.~\ref{fig:enbudget}, the time series of~$\Phi$,~$\Phi_{\rm BH-Inf}$ and~$\Phi_{\rm BH-Disk}$. The injected jet power,~$\dot{\mathcal{E}}_{\rm inj,BH-Inf}$, is very well correlated with the open magnetic flux,~$\Phi_{\rm BH-Inf}$, through the horizon. This is precisely what one expects from the BZ mechanism, here clearly exhibited thanks to our field-line classification scheme. 
We explicitly demonstrate that the BZ correlation,~$\dot{\mathcal{E}}_{\rm inj,BH-Inf} \propto \Phi_{\rm BH-Inf}^2$, is obeyed in Fig.~\ref{fig:bzdemo}.
The proportionality constant can be explained by noting that~$\dot{\mathcal{E}}_{\rm inj,BH-Inf}$ should approach~$f(a) L_{\rm BZ}$, where~$f(a) = 1 + 1.38 (\Omega_{\rm BH} r_g/c)^2 - 9.2 (\Omega_{\rm BH} r_g/c)^4$ is the high-spin correction factor identified by \citet{tchekhovskoy_etal_2010}, as~$\Phi_{\rm BH-Inf}$ tends toward the full fiducial flux~$2 \pi B_0 r_g^2$. This gives~$\dot{\mathcal{E}}_{\rm inj,BH-Inf} / f(a) L_{\rm BZ} = (\Phi_{\rm BH-Inf} / 2 \pi B_0 r_g^2)^2$, as in Fig.~\ref{fig:bzdemo}.

\subsubsection{A toy dissipation model}
\label{sec:enbudgetmodel}
\begin{figure}
    \centering
    \includegraphics[width=\columnwidth]{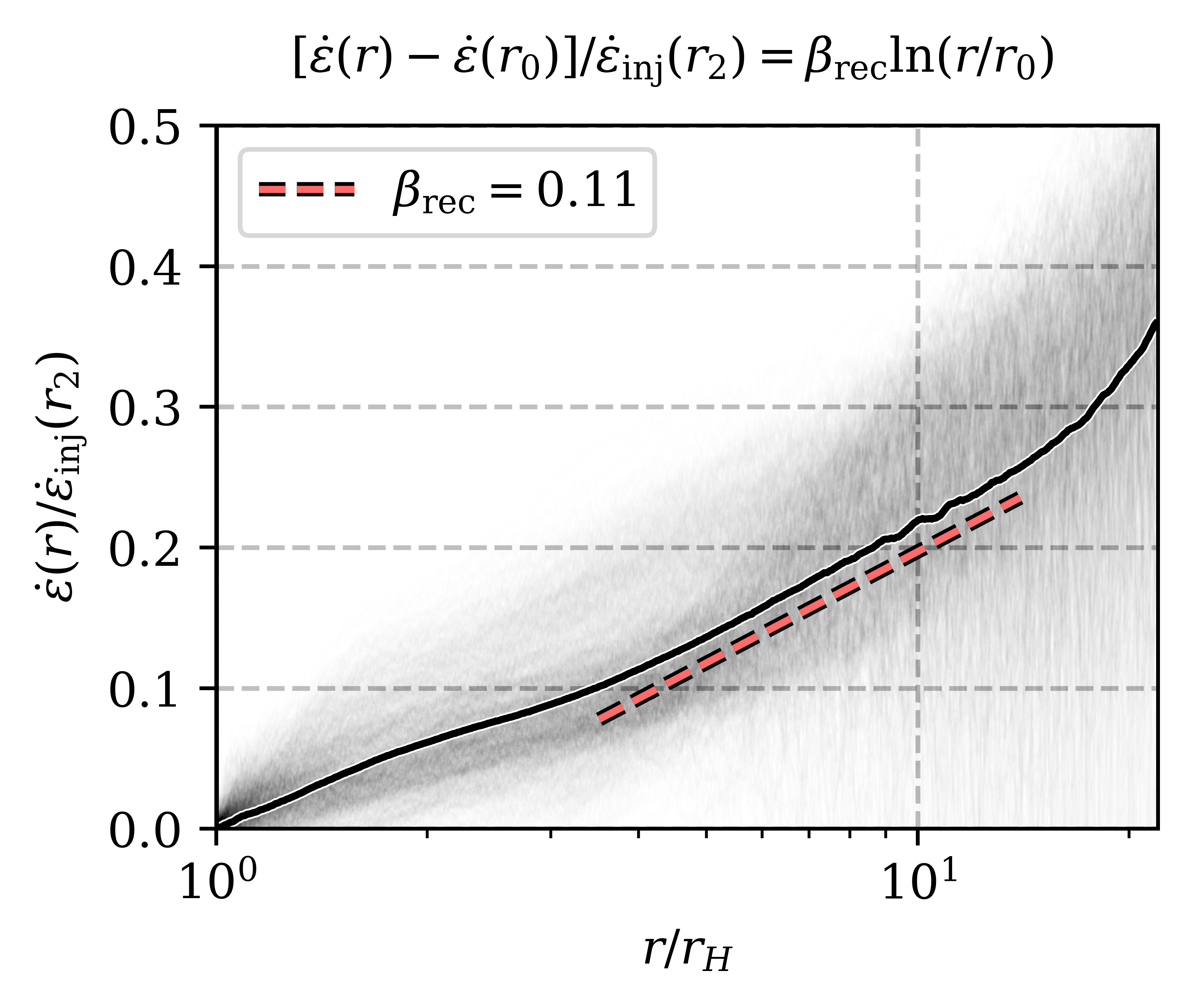}
    \caption{Dissipation profiles computed according to~(\ref{eq:dissprofdef}). Instantaneous profiles are faint black lines; the time-averaged (over~$200 \leq ct / r_g \leq 3200$) profile is a thick black line. We fit~(\ref{eq:toydissrfinal}), with~$\beta_{\rm rec}$ a free parameter, to the time-averaged profile between~$r=r_0=4r_g$ and~$r=16r_g$. The fit is drawn as a dashed red line between~$r=4r_g$ and~$r=16r_g$ with a constant vertical offset for clarity.}
    \label{fig:dissprofs}
\end{figure}
To take a closer look at the physics underlying our dissipation measurements, we present radial dissipation profiles from our simulation in Fig.~\ref{fig:dissprofs}. Each profile is defined with respect to an integration surface similar to that depicted in Fig.~\ref{fig:poyntsurface} as
\begin{align}
    \dissprofvar(r) \equiv \dissprofvar_{\mathrm{inj}}(r) - \int_0^{2\pi} \int_0^{\pi/2} S^r(r, \theta') \sqrt{h} \dif \theta' \dif \phi' \, ,
    \label{eq:dissprofdef}
\end{align}
where
\begin{align}
    \dissprofvar_{\mathrm{inj}}(r) \equiv & \int_0^{2 \pi} \int_{0}^{\pi/2} S^r(\rh, \theta') \sqrt{h} \dif \theta' \dif \phi' \notag \\
    &- \int_{\rh}^{r} \int_0^{2 \pi} S^\theta(r', \theta_{2}) \sqrt{h} \dif r' \dif \phi' \,
    \label{eq:dissprofinj}
\end{align}
is the Poynting flux injected by the BH plus the disk up through radius~$r$.
In the above expressions, as in Fig.~\ref{fig:poyntsurface},~$\vec{S} = c \vec{E} \times \vec{H} / 4 \pi$ and~$\theta_2 = 89^\circ$.

Computing dissipation profiles using~(\ref{eq:dissprofdef}) instead of directly via the~$\vec{J} \cdot \vec{E}$ term of~(\ref{eq:poynt}) misses the contribution from the time-derivative term in Poynting's Theorem. However, this term disappears when we average over an integer number of loop advection cycles. Thus, after averaging, the profiles defined by~(\ref{eq:dissprofdef}) reconstruct the exact~$\vec{J} \cdot \vec{E}$ dissipation.

Understanding the profiles~$\dissprofvar(r)$ presented in Fig.~\ref{fig:dissprofs} is the main goal of this section. To do so, we formulate below a toy dissipation model heavily inspired by that presented by \citet{cerutti_etal_2020} for the case of reconnection in pulsar winds. Within this model, the logarithmic dependence,~$\dissprofvar(r) \propto \ln(r/r_0)$, of the profile follows naturally from magnetic reconnection. The proportionality constant is, moreover, the collisionless relativistic reconnection rate,~$\beta_{\rm rec}$, well known to be of order~$0.1$.

To begin with, we concentrate on dissipation within the innermost current sheet along the jet funnel. Dissipation farther out -- in current sheets along disk-disk field lines -- also occurs, but it does not consume Poynting flux from the jet. Moreover, outer current sheets power very little high-energy radiation (Section~\ref{sec:synthobs}). This is because the local magnetic field strength and accelerated particle energies (Fig.~\ref{fig:lfomega} and its accompanying film) are lower, diminishing the energies of the emerging photons.

As can be seen from Figs.~\ref{fig:lfomega} and~\ref{fig:maglinks} and their accompanying films, reconnection kicks in at a finite distance,~$r_0$, away from the BH, typically around~$4 r_g$ or so. Since this is fairly far from the event horizon, we approximate space as flat throughout the dissipation zone. We therefore use, for the remainder of this section, only the fields~$\vec{B}$ and~$\vec{E}$ and work exclusively in flat orthonormal spherical coordinates (all lower indices with, e.g.,~$\vec{E}^2 = E_r^2 + E_\theta^2 + E_\phi^2$).

In addition, as previously observed in section~\ref{sec:overall}, the poloidal magnetic field far away from the BH is nearly monopolar.
We therefore approximate the radial magnetic field at~$r>r_0 = 4 r_g$ as that of a magnetic monopole:
\begin{align}
    B_r = \pm \btoyfid \frac{r_0^2}{r^2} \, ,
    \label{eq:toybr}
\end{align}
where~$B_{r_0}$ is the radial field strength at~$r = r_0$ and the overall sign changes across reconnection current sheets.
Within the jet funnel, this field rotates at the constant angular velocity~$\Omega = \ombh / 2$ (Fig.~\ref{fig:lfomega}), which can be associated with a light cylinder at~$r \sin \theta = R_{\rm LC} = 2 c / \ombh \simeq 4 \rh / a \simeq 4 r_g$. The azimuthal field produced by this rotation is
\begin{align}
    B_\phi = \mp \frac{r \sin \theta}{R_{\rm LC}} B_r = \mp \btoyfid \frac{r_0^2}{R_{\rm LC}} \frac{\sin \theta}{r} \, .
    \label{eq:toybphi}
\end{align}
For~$r \sin \theta > R_{\rm LC}$, the field is mostly azimuthal~($|B_\phi| > |B_r|$).

We assume that the jet-funnel current sheet is a thin cone-like slab with opening angle~$\theta_{\rm CS}$, where~$\theta_{\rm CS}$, based on Figs.~\ref{fig:lfomega} and~\ref{fig:maglinks} and their accompanying films, is close to~$45^\circ$. The slab extends outwards starting from~$r = r_0$ and has a full angular thickness~$\Delta \theta = \delta / r$ (constant physical thickness~$\delta$). Dissipation causes the profile~$\dissprofvar(r)$ to grow from its initial value,~$\dissprofvar(r_0)$, at the beginning of the current sheet by the amount 
\begin{align}
    \dissprofvar(r) - \dissprofvar(r_0) &= \int_0^{2 \pi} \int_{0}^{\pi/2} \int_{r_0}^{r} \vec{E} \cdot \vec{J} \, r'^2 \sin \theta \dif r' \dif \theta \dif \phi \notag \\ 
    &\simeq 2 \pi \delta \int_{r_0}^{r} \vec{E} \cdot \vec{J} \, r' \sin \theta_{\rm CS} \dif r' \, .
    \label{eq:toydissr}
\end{align}
In the second step, the integral over~$\theta$ only activates across a thin region of thickness~$\Delta \theta = \delta / r'$ centered on~$\theta = \theta_{\rm CS}$.

To evaluate the remaining radial integral in~(\ref{eq:toydissr}), we need to determine the electric field and current density inside the current sheet. Since the condition~$r \sin \theta > R_{\rm LC}$ holds throughout most of the current sheet, where~$\theta = \theta_{\rm CS} \simeq 45^\circ$ and~$r \geq r_0 = 4 r_g \simeq R_{\rm LC}$, the reconnecting magnetic field is mostly azimuthal. This field reverses by an amount~$\Delta B_\phi \sim 2 B_\phi$ across the reconnection layer, requiring a radial current density of
\begin{align}
    J_r \simeq \frac{c \Delta B_\phi}{4 \pi \delta} \sim \frac{c B_\phi}{2 \pi \delta} \, .
    \label{eq:toyjr}
\end{align}
Meanwhile, the reconnection electric field in the heart of the layer is~$E_r = \beta_{\rm rec} B_\phi$, where~$\beta_{\rm rec} \sim 0.1$ is the collisionless relativistic reconnection rate.

Plugging these estimates for~$E_r$ and~$J_r$ into equation~(\ref{eq:toydissr}) gives the dissipation profile
\begin{align}
    \dissprofvar(r) - \dissprofvar(r_0) &\simeq c \beta_{\rm rec} \btoyfid^2 \frac{r_0^4}{R_{\rm LC}^2} \sin^3 \theta_{\rm CS} \int_{r_0}^r \frac{\dif r'}{r'} \notag \\
    &= c \beta_{\rm rec} \btoyfid^2 \frac{r_0^4}{R_{\rm LC}^2} \sin^3 \theta_{\rm CS} \ln \left( \frac{r}{r_0} \right) \, .
    \label{eq:toydissr2}
\end{align}
The thickness~$\delta$ of the current sheet cancels out. To simplify expression~(\ref{eq:toydissr2}), we normalize it by the Poynting power,~$\dissprofvar_{\rm inj}(r_0)$, injected at~$r=r_0$. Since this radius is before substantial dissipation occurs, we can set~$\dissprofvar(r_0)$ to zero in~(\ref{eq:dissprofdef}), which yields 
\begin{align}
    \dissprofvar_{\rm inj}(r_0) &= \frac{c}{4 \pi} \int_0^{2 \pi} \int_0^{\pi/2} \left( \vec{E} \times \vec{B} \right)_r \, r_0^2 \sin \theta \dif \theta \dif \phi \notag \\
    &= \frac{c}{2} \int_0^{\pi/2} B_\phi^2 \, r_0^2 \sin \theta \dif \theta = \frac{c}{3} \btoyfid^2 \frac{r_0^4}{R_{\rm LC}^2} \, .
    \label{eq:toypoyntinj}
\end{align}
Using this result to normalize~(\ref{eq:toydissr2}), we can write
\begin{align}
    \frac{\dissprofvar(r) - \dissprofvar(r_0)}{\dissprofvar_{\rm inj}(r_0)} \simeq 3 \beta_{\rm rec} \sin^3 \theta_{\rm CS} \ln \left( \frac{r}{r_0} \right) \simeq \beta_{\rm rec} \ln \left( \frac{r}{r_0} \right) \, .
    \label{eq:toydissrfinal}
\end{align}
In the final step, we noted, using~$\theta_{\rm CS} \simeq 45^\circ$, that~$3 \sin^3 \theta_{\rm CS} \simeq 1$.

In Fig.~\ref{fig:dissprofs}, we fit~(\ref{eq:toydissrfinal}) to the time-averaged dissipation profiles measured from our simulation. Unlike in our toy model, the measured profiles exhibit some dissipation at radii smaller than~$r_0 = 4 r_g$. This near-horizon dissipation is probably spurious, related to our loop advection prescription that starts to lose realism there. Thus, when fitting to the measured profiles, we consider only radii~$r>4r_g$. In addition, the dissipation profiles all steepen beyond~$r=16r_g$ or so. Such steepening is expected due to dissipation in the outer current sheet, which is barely contained in our simulation. To avoid this steepening influencing the fit, we consider only~$r<16 r_g$.

The measured profiles of Fig.~\ref{fig:dissprofs} are normalized by the injected energy~$\dissprofvar_{\mathrm{inj}}(r_2)$ up to outer radius~$r_2=25r_g$. Poynting's Theorem dictates that~$\dissprofvar_{\rm inj}(r_2) = \dissprofvar(r_2)-\dissprofvar(r_0) + \dissprofvar_{\rm inj}(r_0)$. Plugging this into~(\ref{eq:toydissrfinal}) yields
\begin{align}
    \frac{\dissprofvar(r)-\dissprofvar(r_0)}{\dissprofvar_{\rm inj}(r_2)} = \frac{\beta_{\rm rec} \ln (r/r_0)}{1+\beta_{\rm rec} \ln(r_2/r_0)} \simeq \beta_{\rm rec} \ln \left( \frac{r}{r_0} \right) \, ,
    \label{eq:toydissrnewnorm}
\end{align}
where, in the last step, we used~$\beta_{\rm rec} \sim 0.1$ to estimate~$1+\beta_{\rm rec}\ln(r_2/r_0) \simeq 1.2 \simeq 1$. Thus, normalization by~$\dissprofvar_{\rm inj}(r_2)$ as in Fig.~\ref{fig:dissprofs} does not substantially change the fitted~$\beta_{\rm rec}$ values.

Extrapolating~(\ref{eq:toydissrfinal}) beyond our numerical domain implies that our simulation underestimates the total dissipation. At extremely large distances, our toy model even implies that the injected Poynting flux becomes fully dissipated at~$r = r_{\rm diss} = r_0 \exp{(1/\beta_{\rm rec})} \sim 10^5 r_g$ (between~$10^3 r_g$ and~$10^9 r_g$ assuming~$0.05 < \beta_{\rm rec} < 0.2$). Taking into account reconnection in the outer current sheet would pull in the radius of complete dissipation even closer. However, since the outer current sheet cannot siphon Poynting flux off of the jet, we think that this estimate of~$r_{\rm diss}$ is appropriate for determining the scale where the jet becomes fully dissipated.

However, whether complete dissipation truly occurs at such enormous radii is quite uncertain. At these scales~($10^5 r_g = 5 \, \rm pc$ for a~$10^9 M_\odot$ BH), the jet may have already started interacting with the ambient medium, modifying our predicted profile~(\ref{eq:toydissrfinal}), if not disrupting the jet entirely. Even without an ambient medium, the jet could still become intrinsically unstable to disruption, for example by kink modes.

We conclude, therefore, that our simulation strictly underestimates total dissipation, much of which occurs at larger, unsimulated radii. Indeed, the toy model of this section implies that reconnection along the jet wall provides an efficient mechanism for converting an initially Poynting-flux-dominated flow into particle kinetic energy and radiation. However, whether Poynting flux dissipation continues according to profile~(\ref{eq:toydissrfinal}) all the way to total depletion depends on additional factors (such as the ambient medium and current-driven instabilities) that are beyond the scope of our model. 

\subsubsection{Energy budget summary}
\label{sec:enbudgetsummarize}

We wish to emphasize the following key takeaways from our measurements (Section~\ref{sec:enbudgetmeasure}) and model (Section~\ref{sec:enbudgetmodel}) of the energy budget in our simulation:
\begin{enumerate}
    \item Our simulation dissipates about one third of the Poynting flux injected by the equatorial disk and the BH; it lets the remaining two thirds pass through the box. \label{enum:enbudgetglobal}
    \item Loop inflation leads to open magnetic flux bundles attached to both the BH and the disk. We refer to these bundles, respectively, as the jet and disk wind.
    \item The jet and wind individually transmit roughly two thirds -- equivalent to about~$0.2L_{\rm BZ}$ and~$0.3 L_{\rm BZ}$, respectively -- of their initial Poynting flux out of the simulation. \label{enum:enbudgetbhdisk}
    \item The wind is quasistationary, but the jet flickers.
    \item Jet flickering results from alternating periods of open magnetic flux accumulating on the BH and reconnection eating that flux away. 
    \item We present a model where reconnection on the jet wall yields the radial Poynting flux dissipation profile,~$\beta_{\rm rec} \ln (r / r_0)$, with~$\beta_{\rm rec} \sim 0.1$ the relativistic reconnection rate.
    \item Our model suggests that dissipation would continue at radii beyond our simulated domain. Thus, points \ref{enum:enbudgetglobal} and \ref{enum:enbudgetbhdisk} above must be interpreted as lower bounds on energy dissipation and upper bounds on Poynting flux transmission.
\end{enumerate}
We caution that all of these results apply to the regime of our simulation, involving loops much larger than the BH:~$\lambda = 10 r_g \gg \rh$. In a companion publication \citep{crinquand_etal_inprep}, we illustrate the effects of a smaller loop size,~$\lambda \lesssim \rh$, a regime in which many of the points above no longer hold.

\subsection{Synthetic observables}
\label{sec:synthobs}
\begin{figure}
    \includegraphics{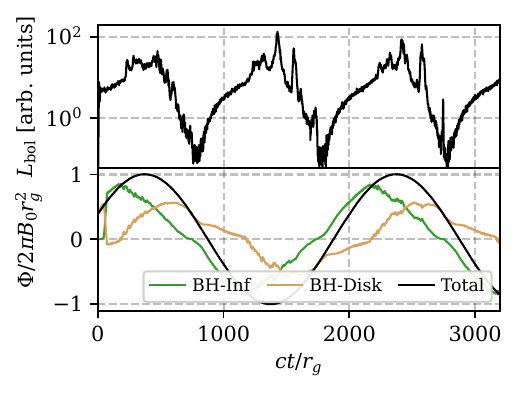}
    \caption{Top: Bolometric luminosity calculated for photons escaping to infinity with energy exceeding~$10 \hbar e B_0 / m_e c$. Here,~$c t / r_g$ corresponds to time of photon emission; not time of photon reception. Light-travel-time effects are thus absent in this lightcurve. Bottom: Instantaneous open (BH-Inf), closed (BH-Disk), and total magnetic flux threading the BH horizon, as in Fig.~\ref{fig:enbudget}.}
    \label{fig:bollc}
\end{figure}

We calculate synchrotron emission from the particles in our simulation and ray-trace \citep[][]{crinquand_etal_2021, crinquand_etal_2022} the resulting photons to observers at infinity. We focus on the high-energy radiation, which is produced by particles accelerated via reconnection. Hence, to avoid contamination from unaccelerated particles (with order-unity Lorentz factors), we only count photons with energy-at-infinity exceeding~$10 \hbar e B_0 / m_e c$. In addition, we ignore photons produced at altitudes lower than~$r \cos \theta = \rh$ above the disk. This excludes spurious near-horizon emission resulting from our \textit{ad hoc} accretion prescription at the equatorial boundary. Figure~\ref{fig:bollc} presents the lightcurve of bolometric luminosity,~$L_{\rm bol}$, of all escaping photons (i.e., neither absorbed by the BH nor by the equatorial disk) as well as the simultaneous horizon magnetic flux calculated per~(\ref{eq:fluxdef}) and~(\ref{eq:fluxdefi}).

The loop advection and ejection cycles, with period~$\lambda / v_0 = 1000 r_g / c$, have a dramatic effect on~$L_{\rm bol}$. Bolometric output peaks near maximum~$|\Phi|$. This corresponds to the moment when reconnection is actively shredding away the open field lines attached to the BH, leading to vigorous particle acceleration. As a result, even though peak brightness periods align with maxima in the total flux,~$|\Phi|$, they follow slightly behind peaks in the open flux,~$|\Phi_{\rm BH-Inf}|$. Magnetic reconnection stochastically ejects the flux in each loop in the form of plasmoids, creating jagged short-timescale features on the falling side of each luminosity peak. Once the innermost loop has been completely ejected~($\Phi \simeq 0$), the luminosity goes through a minimum. From there, it slowly grows again as the next loop is brought into place. This gradual brightening results from reconnection intensifying again as the differential rotation and field strength in the next loop become more pronounced at progressively smaller radii. Loud periods~($|\Phi|$ maxima) and quiet periods~($\Phi \simeq 0$) alternate on the loop-advection period,~$\lambda / v_0$, and exhibit a remarkable brightness contrast of~$10^3$.

\begin{figure*}[t]
    \centering
    \includegraphics[width=\linewidth]{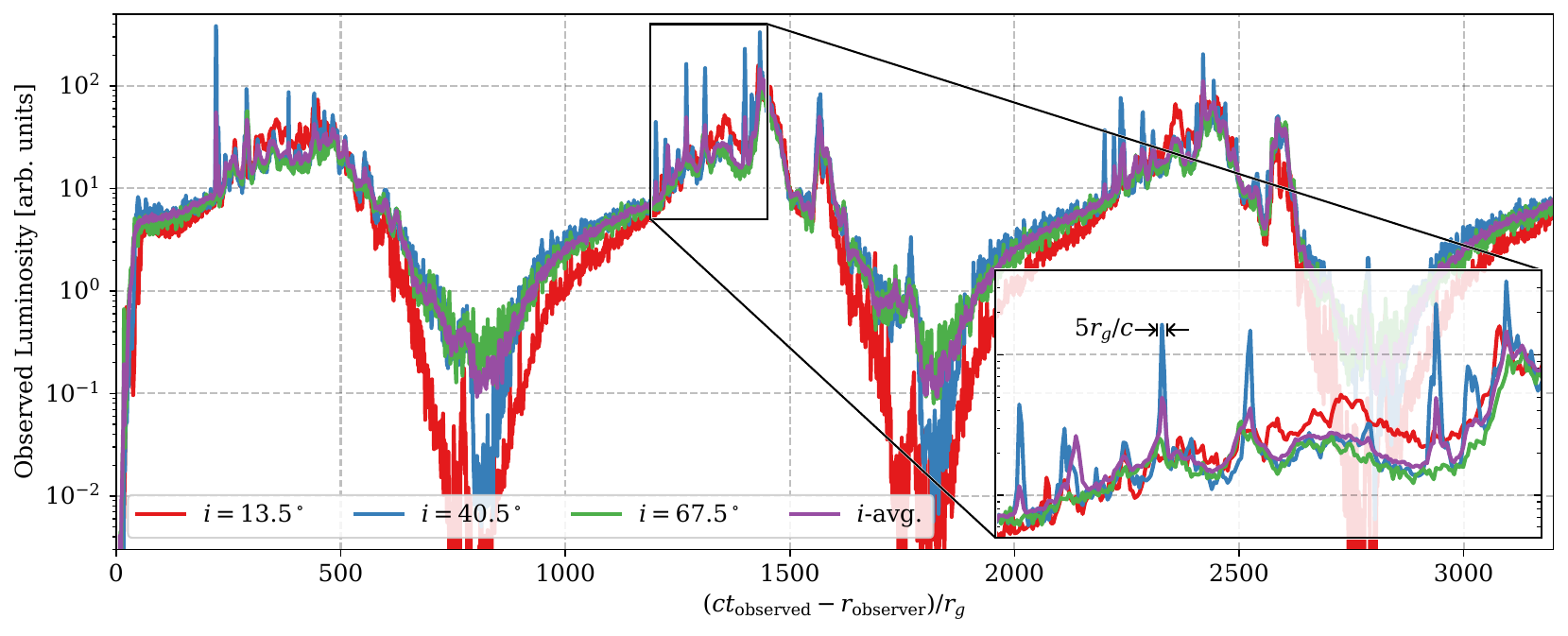}
    \caption{Lightcurves measured by distant observers at different inclinations~$i$. The average lightcurve over all observers between~$i=0^\circ$ and~$i=90^\circ$ is also shown for reference~($i$-avg.). The time-axis is reported in retarded time, correcting for the light-travel-time to the observer (placed an arbitrary distance~$r_{\rm observer}$ away from~$r=0$).}
    \label{fig:synthlc}
\end{figure*}

Besides the bolometric lightcurve, we also present, in Fig.~\ref{fig:synthlc}, lightcurves measured by distant observers at different inclination angles,~$i$, measured from the BH spin axis.
The main features present in the~$L_{\rm bol}$ time series are also present in the lightcurves measured by individual observables: on secular timescales (of order the loop advection period,~$1000 r_g/c$), a brightness contrast of~$10^3$ or more between maximum and minimum activity levels; on short timescales and particularly on the falling segment of the luminosity envelope, abrupt reconnection-driven subflaring. 

Despite overall qualitative agreement with the main features of Fig.~\ref{fig:bollc}, the observer-specific lightcurves of Fig.~\ref{fig:synthlc} are quantitatively more nuanced depending on inclination~$i$.
Observers (e.g.,~$i=40.5^\circ$) looking along the brightest inner current sheet on the jet wall witness faster and more intense variability.
For example, in the inset of Fig.~\ref{fig:synthlc}, the~$i = 40.5^\circ$ lightcurve fluctuates by up to a factor of~$10$ on timescales as short as~$\sim r_g /c$. 

This is a light-travel-time effect, similar to what occurs in blazars: because the emitting plasma travels nearly at the speed of light toward the observer, the radiation is beamed and the arrival times of the photons are compressed, rendering taller and sharper the associated subflares in the lightcurve.
The relativistic motion here probably has both fluid and kinetic origins. On the fluid level, the motion of fast plasmoids, especially small ones, carries the plasma radially outward at relativistic speeds \citep{sironi_etal_2016, petropoulou_etal_2016}. There is also likely a contribution from the kinetic beaming effect, wherein the highest energy particles (and, hence, their radiation) are selectively beamed along the reconnection outflow \citep{cerutti_etal_2012, cerutti_etal_2013, cerutti_etal_2014, mehlhaff_etal_2020}.

\begin{figure}
    \centering
    \includegraphics[width=\columnwidth]{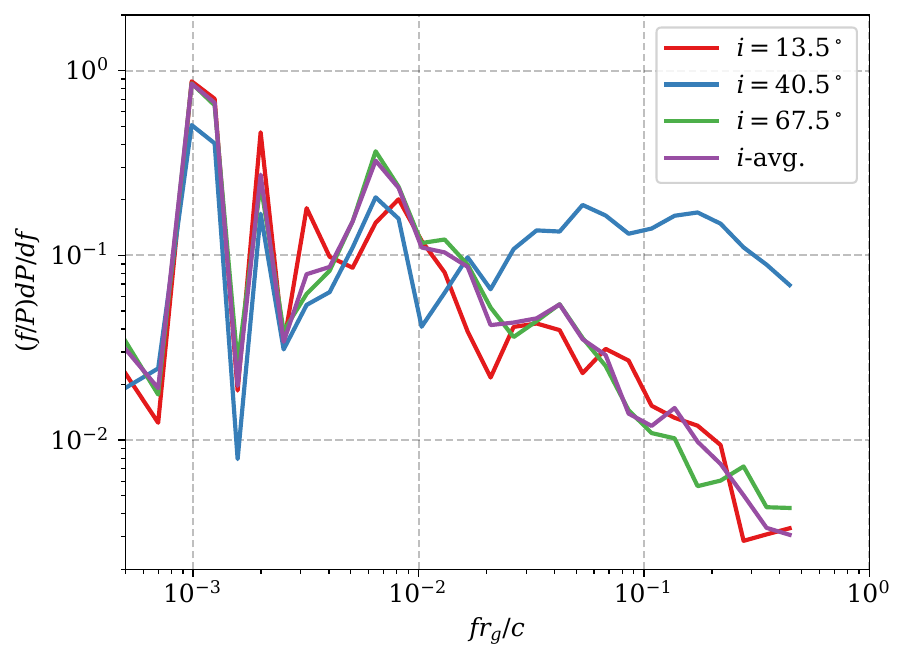}
    \caption{Power spectral densities (PSDs),~$dP/df$, for observers at different inclinations~$i$. These PSDs are calculated from the lightcurves in Fig.~\ref{fig:synthlc}, including the one averaged over all inclination angles. Each PSD is compensated by~$f$, which shows the spectral power per logarithmic frequency interval, and normalized by~$P \equiv \int_0^\infty dP / df \, df$.}
    \label{fig:synthpsd}
\end{figure}
We complement the lightcurves in Fig.~\ref{fig:synthlc} with corresponding power spectral densities (PSDs) in Fig.~\ref{fig:synthpsd}. Each PSD is defined as $d P / d f \equiv (|X(f)| + |X(-f)|)/2$, where~$X(f) \equiv \int_{-\infty}^{\infty} L(t) \exp(-2 \pi i f t) dt$ is the Fourier Transform of the corresponding observer's lightcurve,~$L(t)$. Our secular loop-advection timescale~$1000 r_g / c$ can clearly be seen in Fig.~\ref{fig:synthpsd}: the spectral power for each PSD is concentrated at frequencies~$f \sim 10^{-3} c / r_g$. In contrast to this long-timescale narrow-frequency driving, reconnection produces much faster and broader-band variability, resulting in extended tails in the PSD for each observer. The privileged~$i = 40.5^\circ$ observer that looks directly along the inner reconnection current sheet witnesses the most rapid variability timescales. Hence, the~$i=40.5^\circ$ PSD contains an excess at high frequencies that extends all the way up to the Nyquist limit of~$c / 2 r_g$.

We cannot fully determine the longest variability timescales that reconnection can produce. This is because, at low frequencies in Fig.~\ref{fig:synthpsd}, the narrow-frequency feature associated with the loop-advection timescale,~$\lambda / v_0 = 1000 r_g/c$, overlaps the broadband reconnection-powered variability. A low-frequency cutoff on the power-law PSD produced by reconnection is a natural outcome of the self-similar plasmoid hierarchy \citep{shibata_tanuma_2001, uzdensky_etal_2010}, where the slowest variability timescales are governed by the growth and ejection of the largest plasmoids the current sheet can produce \citep{petropoulou_etal_2016}. Definitively measuring such a cutoff would require pushing the loop-driving feature in our PSDs to lower frequencies, demanding either a slower advection speed,~$v_0$, or a larger loop size,~$\lambda$, and, hence, more expensive simulations. In the absence of such simulations, we infer a conservative bound on the slowest variability frequencies yielded by reconnection. This bound can be deduced from the excess in the~$i=40.5^\circ$ PSD of Fig.~\ref{fig:synthpsd}. Because the excess persists down to frequencies~$f \sim 10^{-2} c/r_g$ and is linked exclusively to anisotropy (beaming and variability compression) associated with reconnection, we infer that reconnection can drive variability on timescales at least as long as~$\sim 100 r_g/c$.

In summary, the high-energy synchrotron observables from our simulations feature:
\begin{enumerate}
    \item A large, secularly driven brightness contrast of at least~$10^3$ (possibly more depending on observer inclination) between bright and quiet phases of the loop advection and ejection cycle.
    \item Peak intensity periods coincident with peak magnetic flux threading the BH.
    \item Rapid subflaring due to reconnection between the BH and equatorial disk. Subflaring begins in the bright period, close to peak magnetic flux on the horizon, and continues during the transition to the quiet phase, as the horizon flux wanes.
    \item Enhanced variability at inclinations matching the opening angle of the innermost current sheet. Subflares at these inclinations can rise by up to an order of magnitude in as little as~$\sim r_g/c$.
    \item Reconnection-driven broadband variability extending across at least two decades in timescales, from~$\sim r_g/c$ up through~$\sim 100 r_g/c$. Costlier simulations could determine if reconnection can yield even slower variability, expanding this range.
\end{enumerate}

\section{Discussion}
\label{sec:discussion}

\subsection{Hard-to-soft X-ray binary state transitions}
\label{sec:xraytrans}
Outbursting BH XRBs pass through a sequence of empirically defined states: first the hard state, where the X-ray spectrum is dominated by nonthermal (power-law) emission peaking in the hard X-rays; next an intermediate state where the overall luminosity is high and the spectrum begins to soften; subsequently the soft state, where a soft,~$\sim 1$ keV quasithermal spectrum dominates the X-ray emission; finally, an eventual return to the hard state \citep{remillard_mcclintock_2006, done_etal_2007}. These X-ray states also have radio counterparts. During the hard state, a compact, steady jet is detected in the radio; in the soft state, the radio emission is consistent with no jet; the hard-to-soft transition is heralded by one or more discrete radio ejections before the jet eventually extinguishes as the binary settles into the soft state \citep{fender_etal_2004, fender_etal_2009, miller-jones_etal_2012, russell_etal_2019}.

We would like to suggest a picture where radio ejections observed during hard-to-soft XRB transitions stem from the launching of magnetic loops as in our simulation: one radio ejection per loop. In this picture, magnetic reconnection links X-ray flaring to radio ejecta by playing dual roles. First, it powers high-energy emission (Section~\ref{sec:synthobs}). Second, it also eats away at the magnetic field lines forming the jet funnel (Section~\ref{sec:enbudgetmeasure}), eventually completely untying their footpoints from the BH, and, as a result, ejecting the magnetic flux originally contained in each coronal loop toward infinity. 

In such a scenario, radio emission is ultimately powered by the Poynting flux injected into the jet by the BH. Observations suggest that this Poynting flux must silently propagate for some distance before dissipating. This is because discrete radio ejections generally light up after the main X-ray transition: already at a considerable distance from the BH. It is only by backtracing the motion of the ejecta across the sky that a launch time nearly coincident with the X-ray transition can be inferred \citep{miller-jones_etal_2012, russell_etal_2019, wood_etal_2021}.

In the context of our model, the movement of the radio ejecta seems to preclude their production by continuous reconnection along the jet wall (Section~\ref{sec:enbudgetmodel}), which would yield a standing structure. Moreover, such reconnection must not entirely consume the available Poynting energy in the jet. Some of it must remain to allow for more sudden and explosive dissipation by different mechanisms farther downstream.

Among the possible downstream dissipation mechanisms are shocks that form as the ejected material interacts with the ambient medium or catches up to slower previously launched ejecta \citep{jamil_etal_2010, malzac_2014}. Alternatively, given that the polarity of each ejected loop alternates, a sequence of ejected loops would, at large scales, produce a magnetically \textit{striped jet}. The Poynting flux in such a jet could be dissipated by direct reconnection between adjacent stripes \citep{drenkhahn_2002, drenkhahn_spruit_2002, lyubarsky_2010, giannios_uzdensky_2019, zhang_giannios_2021}. We return to elaborate this point in section~\ref{sec:stripedjets}.

Besides the change to their X-ray spectrum, BH XRBs also frequently exhibit a few characteristic X-ray timing properties during hard-to-soft state transitions. Often their X-ray variability diminishes and quasi-periodic oscillations (QPOs) at low frequencies~($\lesssim 10 \, \rm Hz$) in their X-ray power spectral densities change form \citep[i.e., type B QPOs may appear;][]{fender_etal_2009, belloni_2010, ingram_motta_2019}. Of these features, the reduction in X-ray variability would be expected from our model. As seen in the lightcurves of Figs.~\ref{fig:bollc} and~\ref{fig:synthlc}, the high-energy radiation powered by magnetic reconnection dims as the flux associated with the innermost loop is ejected away. In such a case, one expects the steadier radiation produced by the underlying accretion disk (not modeled in our setup) to stabilize the overall X-ray output.

All of these remarks seem to produce a rather consistent picture of X-rays and radio ejecta in hard-to-soft XRB transitions being coproduced by the reconnection-mediated ejection of coronal loops. However, there are a few important caveats to keep in mind, particularly with respect to X-ray timing. First, X-ray variability, including low-frequency QPOs, associated with hart-to-soft transitions tends to be characterized at frequencies below~$10$ Hz or so \citep{remillard_mcclintock_2006, belloni_2010}. This is much slower than the reconnection-driven variability we measure in Section~\ref{sec:synthobs}, which occurs from frequencies as high as~$c/r_g \sim 10^4 \, \rm Hz$ down to~$10^{-2} c/r_g \sim 10^2 \, \rm Hz$ (assuming a~$10 M_\odot$ BH mass). As mentioned in Section~\ref{sec:synthobs}, probing whether reconnection can yield variability on even slower timescales requires more expensive simulations. Additionally, given the absence of discrete features in our synthetic PSDs (Fig.~\ref{fig:synthpsd}), QPOs are probably not driven by reconnection. Instead, they are likely related to physics outside our model \citep[e.g., to effects concerning the accretion disk, like Lense-Thirring precession;][]{mario_luigi_1998, fragile_etal_2007, ingram_motta_2019, liska_etal_2021}.

\subsection{The peculiar changing-look AGN 1ES 1927+654}
\label{sec:changinglook}
Some active galactic nuclei (AGNs) undergo rapid changing-look events, where, broadly speaking, their spectral types suddenly change from Type I -- defined by clear Doppler-broadened~($\sim 10^3 \, \rm km\, s^{-1}$) emission lines -- to Type II -- characterized by the absence of broad lines -- or vice versa \citep[e.g.,][and references therein]{tohline_osterbrock_1976, lamassa_etal_2015, macleod_etal_2016, yang_etal_2018}. In December 2017, 1ES 1927+654 became the first AGN to be observed changing look in real time \citep{trakhtenbrot_etal_2019, ricci_etal_2020}. Among the several peculiar aspects of its behavior, perhaps the most striking is that,~$t \simeq 160$ days after its initial optical/UV brightening by a factor of~$10^{1-2}$, the X-ray luminosity dropped by nearly four orders of magnitude, bottoming out at~$t \simeq 200$ days before rebounding, by~$t \simeq 300$ days, to just beyond its pre-flare level. Throughout this dramatic drop-off and rebound, the intraday X-rays were themselves variable by nearly two orders of magnitude \citep{trakhtenbrot_etal_2019, ricci_etal_2020}.

\citet{scepi_etal_2021} \citep[and ][]{laha_etal_2022} speculated that the peculiar X-ray activity of 1ES 1927+654 observed in 2017-2018 could be the result of the destruction and reformation of an X-ray emitting magnetized accretion disk corona. The optical/UV brightening, in this scenario, signals an increase in the accretion rate in the underlying disk, which inwardly advects magnetic flux of opposing polarity to that of the field that threads the disk initially. The fresh magnetic flux first reconnects with the originally present flux, destroying the magnetized corona, and then accumulates in the inner region, reconstructing a corona of opposite magnetic polarity to the first one.

The scenario suggested by \citet{scepi_etal_2021} bears a striking qualitative resemblance to our simulations. One might imagine, for example, that the state of affairs at the beginning of the flare is similar to the beginning of our simulations: the BH is initially threaded by a large magnetic flux, with reconnection on BH-disk field lines powering the variable X-ray emission. The optical/UV brightening signals a dramatic increase in the accretion speed, turning on~$v_0$. The inner loop then gets pushed onto the BH and ejected, causing a drastic dip in the X-ray emission, and the next loop gets brought into place, with a commensurate rebound in the X-ray brightness. All in all, we probe the regions between the first two peak activity states (from~$200$ to~$1500$ in the abscissa of Fig.~\ref{fig:synthlc}). Quantitatively, our simulations yield approximately the right level of secular brightness contrast (upwards of~$10^3$) and stochastic variability (up to a factor of~$10$ for special, current-sheet-aligned observers) to match the respective long- and short-timescale X-ray variability of almost four and two decades seen in the changing-look event.

Caveats here include that no discrete radio ejections were observed to accompany the optical/UV/X-ray activity (\citealt{laha_etal_2022}; though we note that such ejections did recently accompany an increase in X-ray activity in the same object; \citealt{meyer_etal_2025}). Furthermore, our simulations resemble the case of a classic, optically thick, geometrically thin accretion disk \citep{shakura_sunyaev_1973} sandwiched by a magnetized corona. The accretion timescales for such a disk around a supermassive BH are known to be far too long to explain a changing-look timescale of several months \citep{dexter_begelman_2019}. Nevertheless, recent MHD simulations \citep{jacquemin-ide_etal_2024} show evidence that thicker disks can efficiently produce and advect inwards their own magnetic field loops, producing a circumdisk magnetic configuration that is actually not so different from our model.

\subsection{Striped jets}
\label{sec:stripedjets}
In our model, reconnection allows a BH to untie magnetic field lines from its accretion disk and twist them up into a BZ-like jet. In each loop accretion cycle, the resultant jet funnel comprises a magnetic helix of opposing polarity to the preceding one. A series of loop advection and ejection cycles thus produces a corresponding train of alternating-polarity magnetic slabs in the jet. These slabs are initially very elongated along the jet axis. However, as they expand along the collimation profile of the jet, they spread out transversely while maintaining an approximately fixed spacing in the longitudinal direction. The far downstream region of the jet thus becomes magnetically \textit{striped}, with axially thin but laterally wide stripes of reversing magnetic polarity \citep{giannios_uzdensky_2019}: much like a striped pulsar wind \citep{coroniti_1990}.

As the stripe aspect ratio thins, the stripes become increasingly prone to reconnect with one another. If strong inter-stripe reconnection can be triggered, it drives particle acceleration (with commensurate radiation) and bulk jet acceleration in tandem. In this way, magnetic striping could explain a number of key aspects of AGN and gamma-ray-burst jets, including the need for a high jet Lorentz factor, \textit{in situ} particle acceleration far downstream of the central engine, and a broad range of variability timescales \citep{drenkhahn_2002, drenkhahn_spruit_2002, giannios_2006, lyubarsky_2010, giannios_uzdensky_2019, zhang_giannios_2021}.

However, due to the tremendous range of scales covered by BH jets (from the event horizon to beyond the host galaxy for the most extreme AGNs), global numerical striped jet models have so far proven elusive. Computational work has instead focused on key subcomponents of the problem, especially whether accreting magnetic loops can indeed translate into powerful stripes in the jet \citep{parfrey_etal_2015, yuan_etal_2019, mahlmann_etal_2020, chashkina_etal_2021}. Our work complements these preceding efforts by being the first to analyze striped-jet launching using a first-principles GRPIC framework (though see also \citealt{elmellah_etal_2022} and \citealt{elmellah_etal_2023}, similar studies without jet striping). Our simulations thus incorporate completely self-consistent energetics. These enable us to confirm a key finding of previous work: the advection of large~($\lambda \gg r_g$) loops leads to efficient jet activation \citep{parfrey_etal_2015, mahlmann_etal_2020, chashkina_etal_2021}.

In addition, the self-consistent particle acceleration and radiation in our simulations enable us to report a clear observable counterpart to the striped-jet launching mechanism.
Magnetic reconnection, as it launches each loop (i.e., eventual stripe) into the jet funnel, consumes a good fraction of the overall energy budget: even slightly more than the Poynting flux that is transmitted into the jet (since~$\langle \dot{\mathcal{E}}_{\rm E.J} \rangle > \langle \dot{\mathcal{E}}_{\rm out,BH\to\infty} \rangle$ in Tables~\ref{table:edotavg} and~\ref{table:edotavg_open}). This powers efficient nonthermal particle acceleration, leading to rapidly variable high-energy emission accompanying each stripe ejection.

While reconnection creates a potentially observable high-energy counterpart to jet striping, it also dissipates the remaining available Poynting flux in the stripes rather efficiently (Section~\ref{sec:enbudgetmodel}), rendering them progressively more kinetic-energy-dominated. This opens up the possibility that the stripes -- if produced as in our simulation -- may become depleted of their magnetic energy before reaching distances where they could otherwise reconnect efficiently. However, we cannot constrain this possibility with high confidence, since our toy model (Section~\ref{sec:enbudgetmodel}) that we use to extrapolate reconnection-powered dissipation to larger scales ignores other factors that become important there.

Other caveats on the implications of our results in the context of striped jets include our~2D setup, which cannot probe important effects such as the potential disruption of the nascent jet due to nonaxisymmetric current-driven (i.e., kink) instabilities \citep{yuan_etal_2019, mahlmann_etal_2020}. Furthermore, in this preliminary study, we have neither examined retrograde disks, which could lead to even more efficient jet launching \citep{parfrey_etal_2015}, nor the impact of varying loop sizes, especially~$\lambda \lesssim r_g$. We will report the effects of alternate loop sizes in a forthcoming companion publication \citep{crinquand_etal_inprep}.

\subsection{Sgr A* flares}
\label{sec:sgrastar}
The supermassive BH at the center of our Galaxy, Sagittarius A* (Sgr A*), exhibits dramatic variability in the infrared (IR) and X-ray bands \citep{witzel_etal_2018}. The \citet{gravity_etal_2018} recently associated IR flaring with the orbital motion of hot spots close to the BH. \citet{elmellah_etal_2022} and \citet{elmellah_etal_2023} proposed a model for the GRAVITY observations based on GRPIC simulations very similar to ours. Within their model, reconnection occurs on field lines coupling the BH to its accretion disk, just like the innermost current sheet in our simulations. As a result, particles are accelerated and ejected away from the BH in the form of hot plasmoids (flux ropes in 3D). \citet{elmellah_etal_2023} argue that such flux ropes could be the structures observed by GRAVITY as hot spots.

A caveat discussed by \citet{elmellah_etal_2023} is that single flux ropes are ejected too quickly to explain the observed flare duration of~30-$60 \, \rm min$, corresponding to~90-$180r_g/c$ for a BH mass of~$4.1 \times 10^6 M_\odot$. They therefore suggest that the GRAVITY hot spots could be the conglomerate of multiple flux ropes. Individual flux ropes could then explain the subflaring observed on timescales of~$\sim 10 \, \rm min$ \citep{genzel_etal_2003, doddseden_etal_2011}, or about~$30 r_g/c$.

Our results would seem to corroborate the conclusions drawn by \citet{elmellah_etal_2023}. In Section~\ref{sec:synthobs}, we show that reconnection generates broadband variability across a range of timescales: from~$r_g/c$ up to~$100 r_g/c$ (though we cannot rule out still longer timescales). This range emerges from the ensemble of particle acceleration sites and plasmoids contained within the reconnection layer. Thus, both the quantitative timescales measured from our simulation and the expected physical mechanism are largely consistent with the picture of \citet{elmellah_etal_2023}. We note that the absence of strong observed variability as fast as~$r_g/c$ could stem from the viewing-angle effect identified in Section~\ref{sec:synthobs}: only observers looking along the current sheet witness pronounced variability at the fastest~$r_g/c$ timescales. 

\section{Conclusions}
\label{sec:conclusions}
In this paper, we use the \codename{grzeltron}code \citep{parfrey_etal_2019} to present an axisymmetric GRPIC model of a BH fed by its ADC.
This is the first PIC model to include the coupling between a magnetized plasma corona to a BH as well as an accretion prescription.
Our model confirms and extends previous similar numerical studies based on force-free electrodynamics \citep{parfrey_etal_2015, yuan_etal_2019, mahlmann_etal_2020}.
In particular, we confirm the activation of the BZ mechanism for the case when large -- but nevertheless zero-net-flux -- coronal magnetic loops (with diameters~$\lambda \gg r_g$) are fed to a central rapidly spinning BH.
We extend preceding works by employing a first-principles plasma description, which allows us to compute fully self-consistent energy budget diagnostics~(Section~\ref{sec:enbudget}) and radiative signatures (Section~\ref{sec:synthobs}). 

In Section~\ref{sec:enbudget}, we trace the flow of Poynting flux through our simulation, evaluating Poynting's theorem both across our whole simulation domain and along specific field-line bundles. We also present a phenomenological model for the Poynting flux dissipation through magnetic reconnection. Globally, we find that about one third of the Poynting flux injected into our simulation (by the BH plus the disk) is consumed through coronal reconnection. The other two thirds are transmitted through our numerical domain via a BZ jet (for open field lines connected to the BH) and a force-free wind (for open field lines attached to the disk). This~2:1 breakdown between dissipation and transmission also turns out to characterize the fate of the Poynting flux injected separately along jet field lines and disk-wind field lines.

However, we wish to discourage the interpretation of this~2:1 transmission-dissipation partition as a quantitative prediction. Instead, this ratio must be interpreted in light of our analytic model of Section~\ref{sec:enbudgetmodel}. That model reproduces the radial profile of the fraction of dissipated Poynting flux,~$\beta_{\rm rec} \ln (r/r_0)$, from our simulation, where~$\beta_{\rm rec} \simeq 0.1$ is the relativistic magnetic reconnection rate. Extrapolating this formula implies that dissipation should continue at radii beyond our numerical domain, albeit more slowly (scaling logarithmically in~$r$).
At the same time, it is not clear that the profile,~$\beta_{\rm rec} \ln(r/r_0)$, remains accurate all the way out to the critical radius,~$r_{\rm diss} = r_0 \exp{(1/\beta_{\rm rec})}$, where it predicts complete dissipation. A number of factors could intervene before then: the ambient medium could disrupt the force-free jet propagation; the jet could become intrinsically unstable (e.g., to kink modes); or the jet magnetization could decline as the result of bulk acceleration, altering the reconnection rate \textit{in situ}. In view of these uncertain factors, we wish to cast the~2:1 transmission-dissipation ratio witnessed in our simulation simply as an upper bound. Dissipation continues beyond the simulation, scaling initially as~$\beta_{\rm rec} \ln (r/r_0)$ but eventually changing shape due to physics relevant at larger scales.

We wish to underscore that, while the final stages of jet dissipation remain uncertain, our work reveals a mechanism through which a nascent jet can begin efficiently converting Poynting flux to particle kinetic energy and radiation straight from launch. The triggering of some additional dissipation mechanism at larger scales, though certainly possible, is, in principle, not needed. Left to its own devices, jets launched via the inflation of a zero-net-flux magnetic loop, as in our simulation, will completely dissipate their energy by the time they reach~$r_{\rm diss}$.

One practical consequence of these remarks is that we cannot guarantee -- though we also cannot exclude -- that the magnetically striped BZ jet produced by our simulation retains a large reservoir of magnetic energy to be consumed later on \citep[][see discussion in Section~\ref{sec:stripedjets}]{giannios_uzdensky_2019}. However, we do witness a very robust electromagnetic counterpart to striped-jet launch. As magnetic reconnection injects successive stripes into the jet, it also powers particle acceleration and bright high-energy radiation at the jet base. High-energy flaring may thus herald jet striping. We characterize such flaring in the synchrotron channel in Section~\ref{sec:synthobs}.

In particular, in Section~\ref{sec:synthobs}, we reconstruct and analyze high-energy synchrotron lightcurves from our simulation. These are computed using the general relativistic ray-tracing module developed for \codename{grzeltron}by \citet{crinquand_etal_2021} and \citet{crinquand_etal_2022}. Independently of inclination angle, the lightcurves show a dramatic brightness contrast of at least~$10^3$ on slow timescales linked to the loop advection-ejection period. The closer a given loop is accreted in radius, the stronger the magnetic field and the rotational shear across its footpoints become. For our large loop sizes~($\lambda = 10 r_g$), this means that the high-energy radiation from reconnection on the innermost loop, which couples the BH to the accretion disk, dominates that from all the other loops. This radiation peaks just after the maximum in open magnetic flux piercing the horizon, when vigorous reconnection preys on and whittles down the jet funnel. Reconnection along the jet wall is the main source of high-energy coronal output \citep{sridhar_etal_2025}.

The secular brightness modulations on the loop advection and ejection period provide an envelope on top of which is superimposed much faster stochastic variability associated with reconnection. These rapid stochastic fluctuations are more sensitive than the secular envelope to observer inclination. For observers looking along the bright jet-wall reconnection layer, radiation is compressed, akin to what occurs in blazars. This intensifies the stochastic variability component, leading to subflares occurring on timescales as fast as~$r_g/c$ and with brightness contrasts of up to a factor of~$10$.

The superimposed secular and stochastic variations in our simulations of, respectively, a factor of~$10^3$ and~$10$ are on approximately the right levels to explain the secular and stochastic X-ray variability observed in the peculiar changing-look event of the AGN,~1ES~1927+654 \citep{trakhtenbrot_etal_2019}. Thus, our model fits well with the picture advanced by \citet{scepi_etal_2021} \citep[and][]{laha_etal_2022} in which this changing-look event corresponds to a change in polarity of the ADC associated with the AGN. This scenario is discussed in greater detail in section~\ref{sec:changinglook}.

Additionally, we measure (Section~\ref{sec:synthobs}) a range of reconnection-powered variability timescales: from~$r_g/c$ at the shortest up to (at least)~$100r_g/c$ at the longest. These are roughly commensurate with the range of timescales observed in the context of Sgr A* IR flares \citep{genzel_etal_2003, doddseden_etal_2011, gravity_etal_2018}. Thus, our work corroborates the situation envisaged by \citet{elmellah_etal_2023}, where reconnection along field lines coupling a spinning BH to its accretion disk is responsible for the flares. This application is elaborated further in Section~\ref{sec:sgrastar}.

Our work reveals an important connection between coronal activity and jet launch: BZ jet power (Section~\ref{sec:enbudgetmeasure}) and coronal emission (Section~\ref{sec:synthobs}) are coupled through a single reconnection layer. Thus, reconnection-mediated ejection of a coronal loop by a BH might explain radio-jet ejections accompanying X-ray outbursts during XRB hard-to-soft state transitions \citep[][see Section~\ref{sec:xraytrans}]{fender_etal_2004}. Such a situation also naturally gives rise to the reduction in broadband X-ray variability that typically coincides with these transitions \citep{ingram_motta_2019}. As the stochastically fluctuating (Section~\ref{sec:synthobs}) hard X-rays associated with coronal reconnection diminish, the presumably steadier soft X-ray radiation coming from the disk will dominate the lightcurve. However, discrete features (QPOs) seen in X-ray power spectral densities during XRB state transitions are likely produced by some other mechanism, perhaps linked to the accretion disk, that we do not model.

Our work focuses on the regime expected to launch powerful jets: that is, where the accreting magnetic loops are much larger than the central BH. In a companion publication \citep{crinquand_etal_inprep}, we present the complementary and highly qualitatively different limit where the loops are smaller with respect to the BH. Besides the large loop size, a strong assumption of our work is that of axisymmetry. Relaxing this assumption in the future will open up the possibility for diagnosing the impact of nonaxisymmetric modes \citep{yuan_etal_2019, mahlmann_etal_2020} on jet launch. Another extension of our work would be to add polarization to the ray-tracing analysis. This would help constrain our model in light of recent X-ray polarimetry measurements by the IXPE instrument \citep[e.g.,][]{krawczynski_etal_2022, chattopadhyay_etal_2024}. Finally, and perhaps most exciting, our model represents a fundamental link between BH ADCs on the one hand and the jet properties and dissipation that manifest at large scales on the other. Elucidating this link even further via first-principles modeling -- and with the continued support of the complementary force-free and MHD approaches -- is paramount in the quest to reveal the mysteries of BH jets more broadly.

\begin{acknowledgements}
    The authors gratefully acknowledge stimulating discussions with
    Jonathan Ferreira,
    Jeremy Goodman,
    Claire Guépin,
    Gilles Henri,
    Geoffroy Lesur,
    Krzysztof Nalewajko,
    Pierre-Olivier Petrucci,
    Nicolas Scepi,
    Adrien Soudais,
    Dmitri Uzdensky,
    and
    Greg Werner.
    In addition, we would like to thank Adrien Soudais for generously providing the colormap for the left-hand panel of Fig.~\ref{fig:lfomega}.
    This project has received funding from the European Research Council (ERC) under the European Union’s Horizon 2020 research and innovation programme (grant agreement no.\ 863412).
    Computing resources were provided by TGCC under the allocation A0150407669 made by GENCI.
\end{acknowledgements}

\bibliographystyle{aa}
\bibliography{ref}

\begin{thebibliography}{114}
\expandafter\ifx\csname natexlab\endcsname\relax\def\natexlab#1{#1}\fi

\bibitem[{{Aly}(1984)}]{aly_1984}
{Aly}, J.~J. 1984, \apj, 283, 349

\bibitem[{{Aly}(1991)}]{aly_1991}
{Aly}, J.~J. 1991, \apjl, 375, L61

\bibitem[{{Barnes} \& {Sturrock}(1972)}]{barnes_sturrock_1972}
{Barnes}, C.~W. \& {Sturrock}, P.~A. 1972, \apj, 174, 659

\bibitem[{{Belloni}(2010)}]{belloni_2010}
{Belloni}, T.~M. 2010, in Lecture Notes in Physics, Berlin Springer Verlag, ed.
  T.~{Belloni}, Vol. 794, 53

\bibitem[{{Beloborodov}(2017)}]{beloborodov_2017}
{Beloborodov}, A.~M. 2017, \apj, 850, 141

\bibitem[{{Berenger}(1994)}]{berenger_1994}
{Berenger}, J.-P. 1994, Journal of Computational Physics, 114, 185

\bibitem[{{Birdsall} \& {Langdon}(1991)}]{birdsall_langdon_1991}
{Birdsall}, C.~K. \& {Langdon}, A.~B. 1991, {Plasma Physics via Computer
  Simulation}

\bibitem[{{Bisnovatyi-Kogan} \&
  {Blinnikov}(1976)}]{bisnovatyiKogan_blinnikov_1976}
{Bisnovatyi-Kogan}, G.~S. \& {Blinnikov}, S.~I. 1976, Soviet Astronomy Letters,
  2, 191

\bibitem[{{Blackman} \& {Field}(1994)}]{blackman_field_1994}
{Blackman}, E.~G. \& {Field}, G.~B. 1994, \prl, 72, 494

\bibitem[{{Blandford}(1976)}]{blandford_1976}
{Blandford}, R.~D. 1976, \mnras, 176, 465

\bibitem[{{Blandford} \& {Payne}(1982)}]{blandford_payne_1982}
{Blandford}, R.~D. \& {Payne}, D.~G. 1982, \mnras, 199, 883

\bibitem[{{Blandford} \& {Znajek}(1977)}]{blandford_znajek_1977}
{Blandford}, R.~D. \& {Znajek}, R.~L. 1977, \mnras, 179, 433

\bibitem[{{Cerutti} {et~al.}(2015){Cerutti}, {Philippov}, {Parfrey}, \&
  {Spitkovsky}}]{cerutti_etal_2015}
{Cerutti}, B., {Philippov}, A., {Parfrey}, K., \& {Spitkovsky}, A. 2015,
  \mnras, 448, 606

\bibitem[{{Cerutti} {et~al.}(2020){Cerutti}, {Philippov}, \&
  {Dubus}}]{cerutti_etal_2020}
{Cerutti}, B., {Philippov}, A.~A., \& {Dubus}, G. 2020, \aap, 642, A204

\bibitem[{{Cerutti} {et~al.}(2012){Cerutti}, {Werner}, {Uzdensky}, \&
  {Begelman}}]{cerutti_etal_2012}
{Cerutti}, B., {Werner}, G.~R., {Uzdensky}, D.~A., \& {Begelman}, M.~C. 2012,
  \apjl, 754, L33

\bibitem[{{Cerutti} {et~al.}(2013){Cerutti}, {Werner}, {Uzdensky}, \&
  {Begelman}}]{cerutti_etal_2013}
{Cerutti}, B., {Werner}, G.~R., {Uzdensky}, D.~A., \& {Begelman}, M.~C. 2013,
  \apj, 770, 147

\bibitem[{{Cerutti} {et~al.}(2014){Cerutti}, {Werner}, {Uzdensky}, \&
  {Begelman}}]{cerutti_etal_2014}
{Cerutti}, B., {Werner}, G.~R., {Uzdensky}, D.~A., \& {Begelman}, M.~C. 2014,
  \apj, 782, 104

\bibitem[{{Chashkina} {et~al.}(2021){Chashkina}, {Bromberg}, \&
  {Levinson}}]{chashkina_etal_2021}
{Chashkina}, A., {Bromberg}, O., \& {Levinson}, A. 2021, \mnras, 508, 1241

\bibitem[{{Chattopadhyay} {et~al.}(2024){Chattopadhyay}, {Kumar}, {Rao},
  {Bhargava}, {Vadawale}, {Ratheesh}, {Dewangan}, {Bhattacharya}, {Mithun}, \&
  {Bhalerao}}]{chattopadhyay_etal_2024}
{Chattopadhyay}, T., {Kumar}, A., {Rao}, A.~R., {et~al.} 2024, \apjl, 960, L2

\bibitem[{{Coroniti}(1990)}]{coroniti_1990}
{Coroniti}, F.~V. 1990, \apj, 349, 538

\bibitem[{{Crinquand}(et al. in prep.)}]{crinquand_etal_inprep}
{Crinquand}, B. et al. in prep.

\bibitem[{{Crinquand} {et~al.}(2021){Crinquand}, {Cerutti}, {Dubus}, {Parfrey},
  \& {Philippov}}]{crinquand_etal_2021}
{Crinquand}, B., {Cerutti}, B., {Dubus}, G., {Parfrey}, K., \& {Philippov}, A.
  2021, \aap, 650, A163

\bibitem[{{Crinquand} {et~al.}(2022){Crinquand}, {Cerutti}, {Dubus}, {Parfrey},
  \& {Philippov}}]{crinquand_etal_2022}
{Crinquand}, B., {Cerutti}, B., {Dubus}, G., {Parfrey}, K., \& {Philippov}, A.
  2022, \prl, 129, 205101

\bibitem[{{Crinquand} {et~al.}(2020){Crinquand}, {Cerutti}, {Philippov},
  {Parfrey}, \& {Dubus}}]{crinquand_etal_2020}
{Crinquand}, B., {Cerutti}, B., {Philippov}, A., {Parfrey}, K., \& {Dubus}, G.
  2020, \prl, 124, 145101

\bibitem[{{de Gouveia dal Pino} \&
  {Lazarian}(2005)}]{deGouveiaDalPino_lazarian_2005}
{de Gouveia dal Pino}, E.~M. \& {Lazarian}, A. 2005, \aap, 441, 845

\bibitem[{{Dexter} \& {Begelman}(2019)}]{dexter_begelman_2019}
{Dexter}, J. \& {Begelman}, M.~C. 2019, \mnras, 483, L17

\bibitem[{{Di Matteo}(1998)}]{diMatteo_1998}
{Di Matteo}, T. 1998, \mnras, 299, L15

\bibitem[{{Dodds-Eden} {et~al.}(2011){Dodds-Eden}, {Gillessen}, {Fritz},
  {Eisenhauer}, {Trippe}, {Genzel}, {Ott}, {Bartko}, {Pfuhl}, {Bower},
  {Goldwurm}, {Porquet}, {Trap}, \& {Yusef-Zadeh}}]{doddseden_etal_2011}
{Dodds-Eden}, K., {Gillessen}, S., {Fritz}, T.~K., {et~al.} 2011, \apj, 728, 37

\bibitem[{{Done} {et~al.}(2007){Done}, {Gierli{\'n}ski}, \&
  {Kubota}}]{done_etal_2007}
{Done}, C., {Gierli{\'n}ski}, M., \& {Kubota}, A. 2007, \aapr, 15, 1

\bibitem[{{Drenkhahn}(2002)}]{drenkhahn_2002}
{Drenkhahn}, G. 2002, \aap, 387, 714

\bibitem[{{Drenkhahn} \& {Spruit}(2002)}]{drenkhahn_spruit_2002}
{Drenkhahn}, G. \& {Spruit}, H.~C. 2002, \aap, 391, 1141

\bibitem[{{El Mellah} {et~al.}(2023){El Mellah}, {Cerutti}, \&
  {Crinquand}}]{elmellah_etal_2023}
{El Mellah}, I., {Cerutti}, B., \& {Crinquand}, B. 2023, \aap, 677, A67

\bibitem[{{El Mellah} {et~al.}(2022){El Mellah}, {Cerutti}, {Crinquand}, \&
  {Parfrey}}]{elmellah_etal_2022}
{El Mellah}, I., {Cerutti}, B., {Crinquand}, B., \& {Parfrey}, K. 2022, \aap,
  663, A169

\bibitem[{{Fender} {et~al.}(2004){Fender}, {Belloni}, \&
  {Gallo}}]{fender_etal_2004}
{Fender}, R.~P., {Belloni}, T.~M., \& {Gallo}, E. 2004, \mnras, 355, 1105

\bibitem[{{Fender} {et~al.}(2009){Fender}, {Homan}, \&
  {Belloni}}]{fender_etal_2009}
{Fender}, R.~P., {Homan}, J., \& {Belloni}, T.~M. 2009, \mnras, 396, 1370

\bibitem[{{Fragile} {et~al.}(2007){Fragile}, {Blaes}, {Anninos}, \&
  {Salmonson}}]{fragile_etal_2007}
{Fragile}, P.~C., {Blaes}, O.~M., {Anninos}, P., \& {Salmonson}, J.~D. 2007,
  \apj, 668, 417

\bibitem[{{Galeev} {et~al.}(1979){Galeev}, {Rosner}, \&
  {Vaiana}}]{galeev_etal_1979}
{Galeev}, A.~A., {Rosner}, R., \& {Vaiana}, G.~S. 1979, \apj, 229, 318

\bibitem[{{Genzel} {et~al.}(2003){Genzel}, {Sch{\"o}del}, {Ott}, {Eckart},
  {Alexander}, {Lacombe}, {Rouan}, \& {Aschenbach}}]{genzel_etal_2003}
{Genzel}, R., {Sch{\"o}del}, R., {Ott}, T., {et~al.} 2003, \nat, 425, 934

\bibitem[{{Giannios}(2006)}]{giannios_2006}
{Giannios}, D. 2006, \aap, 457, 763

\bibitem[{{Giannios} \& {Uzdensky}(2019)}]{giannios_uzdensky_2019}
{Giannios}, D. \& {Uzdensky}, D.~A. 2019, \mnras, 484, 1378

\bibitem[{{Goodman}(2003)}]{goodman_2003}
{Goodman}, J. 2003, \mnras, 339, 937

\bibitem[{{Goodman} \& {Uzdensky}(2008)}]{goodman_uzdensky_2008}
{Goodman}, J. \& {Uzdensky}, D. 2008, \apj, 688, 555

\bibitem[{{GRAVITY Collaboration} {et~al.}(2018){GRAVITY Collaboration},
  {Abuter}, {Amorim}, {Baub{\"o}ck}, {Berger}, {Bonnet}, {Brandner},
  {Cl{\'e}net}, {Coud{\'e} Du Foresto}, {de Zeeuw}, {Deen}, {Dexter}, {Duvert},
  {Eckart}, {Eisenhauer}, {F{\"o}rster Schreiber}, {Garcia}, {Gao}, {Gendron},
  {Genzel}, {Gillessen}, {Guajardo}, {Habibi}, {Haubois}, {Henning}, {Hippler},
  {Horrobin}, {Huber}, {Jim{\'e}nez-Rosales}, {Jocou}, {Kervella}, {Lacour},
  {Lapeyr{\`e}re}, {Lazareff}, {Le Bouquin}, {L{\'e}na}, {Lippa}, {Ott},
  {Panduro}, {Paumard}, {Perraut}, {Perrin}, {Pfuhl}, {Plewa}, {Rabien},
  {Rodr{\'\i}guez-Coira}, {Rousset}, {Sternberg}, {Straub}, {Straubmeier},
  {Sturm}, {Tacconi}, {Vincent}, {von Fellenberg}, {Waisberg}, {Widmann},
  {Wieprecht}, {Wiezorrek}, {Woillez}, \& {Yazici}}]{gravity_etal_2018}
{GRAVITY Collaboration}, {Abuter}, R., {Amorim}, A., {et~al.} 2018, \aap, 618,
  L10

\bibitem[{{Gro{\v{s}}elj} {et~al.}(2024){Gro{\v{s}}elj}, {Hakobyan},
  {Beloborodov}, {Sironi}, \& {Philippov}}]{groselj_etal_2024}
{Gro{\v{s}}elj}, D., {Hakobyan}, H., {Beloborodov}, A.~M., {Sironi}, L., \&
  {Philippov}, A. 2024, \prl, 132, 085202

\bibitem[{{Gupta} {et~al.}(2024){Gupta}, {Sridhar}, \&
  {Sironi}}]{gupta_etal_2024}
{Gupta}, S., {Sridhar}, N., \& {Sironi}, L. 2024, \mnras, 527, 6065

\bibitem[{{Heyvaerts} \& {Priest}(1989)}]{heyvaerts_priest_1989}
{Heyvaerts}, J.~F. \& {Priest}, E.~R. 1989, \aap, 216, 230

\bibitem[{{Ingram} \& {Motta}(2019)}]{ingram_motta_2019}
{Ingram}, A.~R. \& {Motta}, S.~E. 2019, \nar, 85, 101524

\bibitem[{{Jackson}(1975)}]{jackson_1975}
{Jackson}, J.~D. 1975, {Classical electrodynamics}

\bibitem[{{Jacquemin-Ide} {et~al.}(2021){Jacquemin-Ide}, {Lesur}, \&
  {Ferreira}}]{jacquemin-ide_etal_2021}
{Jacquemin-Ide}, J., {Lesur}, G., \& {Ferreira}, J. 2021, \aap, 647, A192

\bibitem[{{Jacquemin-Ide} {et~al.}(2024){Jacquemin-Ide}, {Rincon},
  {Tchekhovskoy}, \& {Liska}}]{jacquemin-ide_etal_2024}
{Jacquemin-Ide}, J., {Rincon}, F., {Tchekhovskoy}, A., \& {Liska}, M. 2024,
  \mnras, 532, 1522

\bibitem[{{Jamil} {et~al.}(2010){Jamil}, {Fender}, \&
  {Kaiser}}]{jamil_etal_2010}
{Jamil}, O., {Fender}, R.~P., \& {Kaiser}, C.~R. 2010, \mnras, 401, 394

\bibitem[{{Kadowaki} {et~al.}(2015){Kadowaki}, {de Gouveia Dal Pino}, \&
  {Singh}}]{kadowaki_etal_2015}
{Kadowaki}, L.~H.~S., {de Gouveia Dal Pino}, E.~M., \& {Singh}, C.~B. 2015,
  \apj, 802, 113

\bibitem[{{Khiali} {et~al.}(2015){Khiali}, {de Gouveia Dal Pino}, \& {del
  Valle}}]{khiali_etal_2015}
{Khiali}, B., {de Gouveia Dal Pino}, E.~M., \& {del Valle}, M.~V. 2015, \mnras,
  449, 34

\bibitem[{{Komissarov}(2004)}]{komissarov_2004}
{Komissarov}, S.~S. 2004, \mnras, 350, 427

\bibitem[{{Konigl}(1989)}]{konigl_1989}
{Konigl}, A. 1989, \apj, 342, 208

\bibitem[{{Krawczynski} {et~al.}(2022){Krawczynski}, {Muleri}, {Dov{\v{c}}iak},
  {Veledina}, {Rodriguez Cavero}, {Svoboda}, {Ingram}, {Matt}, {Garcia},
  {Loktev}, {Negro}, {Poutanen}, {Kitaguchi}, {Podgorn{\'y}}, {Rankin},
  {Zhang}, {Berdyugin}, {Berdyugina}, {Bianchi}, {Blinov}, {Capitanio}, {Di
  Lalla}, {Draghis}, {Fabiani}, {Kagitani}, {Kravtsov}, {Kiehlmann},
  {Latronico}, {Lutovinov}, {Mandarakas}, {Marin}, {Marinucci}, {Miller},
  {Mizuno}, {Molkov}, {Omodei}, {Petrucci}, {Ratheesh}, {Sakanoi}, {Semena},
  {Skalidis}, {Soffitta}, {Tennant}, {Thalhammer}, {Tombesi}, {Weisskopf},
  {Wilms}, {Zhang}, {Agudo}, {Antonelli}, {Bachetti}, {Baldini}, {Baumgartner},
  {Bellazzini}, {Bongiorno}, {Bonino}, {Brez}, {Bucciantini}, {Castellano},
  {Cavazzuti}, {Ciprini}, {Costa}, {De Rosa}, {Del Monte}, {Di Gesu}, {Di
  Marco}, {Donnarumma}, {Doroshenko}, {Ehlert}, {Enoto}, {Evangelista},
  {Ferrazzoli}, {Gunji}, {Hayashida}, {Heyl}, {Iwakiri}, {Jorstad}, {Karas},
  {Kolodziejczak}, {La Monaca}, {Liodakis}, {Maldera}, {Manfreda}, {Marscher},
  {Marshall}, {Mitsuishi}, {Ng}, {O{\textquoteright}Dell}, {Oppedisano},
  {Papitto}, {Pavlov}, {Peirson}, {Perri}, {Pesce-Rollins}, {Pilia},
  {Possenti}, {Puccetti}, {Ramsey}, {Romani}, {Sgr{\`o}}, {Slane}, {Spandre},
  {Tamagawa}, {Tavecchio}, {Taverna}, {Tawara}, {Thomas}, {Trois}, {Tsygankov},
  {Turolla}, {Vink}, {Wu}, {Xie}, \& {Zane}}]{krawczynski_etal_2022}
{Krawczynski}, H., {Muleri}, F., {Dov{\v{c}}iak}, M., {et~al.} 2022, Science,
  378, 650

\bibitem[{{Laha} {et~al.}(2022){Laha}, {Meyer}, {Roychowdhury}, {Becerra
  Gonzalez}, {Acosta-Pulido}, {Thapa}, {Ghosh}, {Behar}, {Gallo}, {Kriss},
  {Panessa}, {Bianchi}, {La Franca}, {Scepi}, {Begelman}, {Longinotti},
  {Lusso}, {Oates}, {Nicholl}, \& {Cenko}}]{laha_etal_2022}
{Laha}, S., {Meyer}, E., {Roychowdhury}, A., {et~al.} 2022, \apj, 931, 5

\bibitem[{{LaMassa} {et~al.}(2015){LaMassa}, {Cales}, {Moran}, {Myers},
  {Richards}, {Eracleous}, {Heckman}, {Gallo}, \& {Urry}}]{lamassa_etal_2015}
{LaMassa}, S.~M., {Cales}, S., {Moran}, E.~C., {et~al.} 2015, \apj, 800, 144

\bibitem[{{Liang} \& {Price}(1977)}]{liang_price_1977}
{Liang}, E.~P.~T. \& {Price}, R.~H. 1977, \apj, 218, 247

\bibitem[{{Liska} {et~al.}(2021){Liska}, {Hesp}, {Tchekhovskoy}, {Ingram}, {van
  der Klis}, {Markoff}, \& {Van Moer}}]{liska_etal_2021}
{Liska}, M., {Hesp}, C., {Tchekhovskoy}, A., {et~al.} 2021, \mnras, 507, 983

\bibitem[{{Liska} {et~al.}(2019){Liska}, {Tchekhovskoy}, {Ingram}, \& {van der
  Klis}}]{liska_etal_2019}
{Liska}, M., {Tchekhovskoy}, A., {Ingram}, A., \& {van der Klis}, M. 2019,
  \mnras, 487, 550

\bibitem[{{Liska} {et~al.}(2020){Liska}, {Tchekhovskoy}, \&
  {Quataert}}]{liska_etal_2020}
{Liska}, M., {Tchekhovskoy}, A., \& {Quataert}, E. 2020, \mnras, 494, 3656

\bibitem[{{Liu} {et~al.}(2002){Liu}, {Mineshige}, \& {Shibata}}]{liu_etal_2002}
{Liu}, B.~F., {Mineshige}, S., \& {Shibata}, K. 2002, \apjl, 572, L173

\bibitem[{{Lubow} {et~al.}(1994){Lubow}, {Papaloizou}, \&
  {Pringle}}]{lubow_etal_1994}
{Lubow}, S.~H., {Papaloizou}, J.~C.~B., \& {Pringle}, J.~E. 1994, \mnras, 267,
  235

\bibitem[{{Lynden-Bell} \& {Boily}(1994)}]{lynden-bell_boily_1994}
{Lynden-Bell}, D. \& {Boily}, C. 1994, \mnras, 267, 146

\bibitem[{{Lyubarsky}(2010)}]{lyubarsky_2010}
{Lyubarsky}, Y. 2010, \apjl, 725, L234

\bibitem[{{Lyubarsky}(2005)}]{lyubarsky_2005}
{Lyubarsky}, Y.~E. 2005, \mnras, 358, 113

\bibitem[{{Lyutikov} \& {Uzdensky}(2003)}]{lyutikov_uzdensky_2003}
{Lyutikov}, M. \& {Uzdensky}, D. 2003, \apj, 589, 893

\bibitem[{{MacLeod} {et~al.}(2016){MacLeod}, {Ross}, {Lawrence}, {Goad},
  {Horne}, {Burgett}, {Chambers}, {Flewelling}, {Hodapp}, {Kaiser}, {Magnier},
  {Wainscoat}, \& {Waters}}]{macleod_etal_2016}
{MacLeod}, C.~L., {Ross}, N.~P., {Lawrence}, A., {et~al.} 2016, \mnras, 457,
  389

\bibitem[{{Mahlmann} {et~al.}(2020){Mahlmann}, {Levinson}, \&
  {Aloy}}]{mahlmann_etal_2020}
{Mahlmann}, J.~F., {Levinson}, A., \& {Aloy}, M.~A. 2020, \mnras, 494, 4203

\bibitem[{{Malzac}(2014)}]{malzac_2014}
{Malzac}, J. 2014, \mnras, 443, 299

\bibitem[{{Mehlhaff} {et~al.}(2024){Mehlhaff}, {Werner}, {Cerutti}, {Uzdensky},
  \& {Begelman}}]{mehlhaff_etal_2024}
{Mehlhaff}, J., {Werner}, G., {Cerutti}, B., {Uzdensky}, D., \& {Begelman}, M.
  2024, \mnras, 527, 11587

\bibitem[{{Mehlhaff} {et~al.}(2020){Mehlhaff}, {Werner}, {Uzdensky}, \&
  {Begelman}}]{mehlhaff_etal_2020}
{Mehlhaff}, J.~M., {Werner}, G.~R., {Uzdensky}, D.~A., \& {Begelman}, M.~C.
  2020, \mnras, 498, 799

\bibitem[{{Mehlhaff} {et~al.}(2021){Mehlhaff}, {Werner}, {Uzdensky}, \&
  {Begelman}}]{mehlhaff_etal_2021}
{Mehlhaff}, J.~M., {Werner}, G.~R., {Uzdensky}, D.~A., \& {Begelman}, M.~C.
  2021, \mnras, 508, 4532

\bibitem[{{Meyer} {et~al.}(2025){Meyer}, {Laha}, {Shuvo}, {Roychowdhury},
  {Green}, {Rhodes}, {Hankla}, {Philippov}, {Mbarek}, {laor}, {Begelman},
  {Sadaula}, {Ghosh}, {Bruni}, {Panessa}, {Guainazzi}, {Behar}, {Masterson},
  {Zhang}, {Yang}, {Gurwell}, {Keating}, {Williams-Baldwin}, {Bray},
  {Bempong-Manful}, {Wrigley}, {Bianchi}, {Ricci}, {La Franca}, {Kara},
  {Georganopoulos}, {Oates}, {Nicholl}, {Pal}, \& {Cenko}}]{meyer_etal_2025}
{Meyer}, E.~T., {Laha}, S., {Shuvo}, O.~I., {et~al.} 2025, \apjl, 979, L2

\bibitem[{{Miller-Jones} {et~al.}(2012){Miller-Jones}, {Sivakoff},
  {Altamirano}, {Coriat}, {Corbel}, {Dhawan}, {Krimm}, {Remillard}, {Rupen},
  {Russell}, {Fender}, {Heinz}, {K{\"o}rding}, {Maitra}, {Markoff}, {Migliari},
  {Sarazin}, \& {Tudose}}]{miller-jones_etal_2012}
{Miller-Jones}, J.~C.~A., {Sivakoff}, G.~R., {Altamirano}, D., {et~al.} 2012,
  \mnras, 421, 468

\bibitem[{{Musoke} {et~al.}(2023){Musoke}, {Liska}, {Porth}, {van der Klis}, \&
  {Ingram}}]{musoke_etal_2023}
{Musoke}, G., {Liska}, M., {Porth}, O., {van der Klis}, M., \& {Ingram}, A.
  2023, \mnras, 518, 1656

\bibitem[{{N{\"a}ttil{\"a}}(2024)}]{nattila_2024}
{N{\"a}ttil{\"a}}, J. 2024, Nature Communications, 15, 7026

\bibitem[{{Novikov} \& {Thorne}(1973)}]{novikov_thorne_1973}
{Novikov}, I.~D. \& {Thorne}, K.~S. 1973, in Black Holes (Les Astres Occlus),
  ed. C.~{Dewitt} \& B.~S. {Dewitt}, 343--450

\bibitem[{{Parfrey} {et~al.}(2015){Parfrey}, {Giannios}, \&
  {Beloborodov}}]{parfrey_etal_2015}
{Parfrey}, K., {Giannios}, D., \& {Beloborodov}, A.~M. 2015, \mnras, 446, L61

\bibitem[{{Parfrey} {et~al.}(2019){Parfrey}, {Philippov}, \&
  {Cerutti}}]{parfrey_etal_2019}
{Parfrey}, K., {Philippov}, A., \& {Cerutti}, B. 2019, \prl, 122, 035101

\bibitem[{{Petropoulou} {et~al.}(2016){Petropoulou}, {Giannios}, \&
  {Sironi}}]{petropoulou_etal_2016}
{Petropoulou}, M., {Giannios}, D., \& {Sironi}, L. 2016, \mnras, 462, 3325

\bibitem[{{Remillard} \& {McClintock}(2006)}]{remillard_mcclintock_2006}
{Remillard}, R.~A. \& {McClintock}, J.~E. 2006, \araa, 44, 49

\bibitem[{{Ricci} {et~al.}(2020){Ricci}, {Kara}, {Loewenstein}, {Trakhtenbrot},
  {Arcavi}, {Remillard}, {Fabian}, {Gendreau}, {Arzoumanian}, {Li}, {Ho},
  {MacLeod}, {Cackett}, {Altamirano}, {Gandhi}, {Kosec}, {Pasham}, {Steiner},
  \& {Chan}}]{ricci_etal_2020}
{Ricci}, C., {Kara}, E., {Loewenstein}, M., {et~al.} 2020, \apjl, 898, L1

\bibitem[{{Russell} {et~al.}(2019){Russell}, {Tetarenko}, {Miller-Jones},
  {Sivakoff}, {Parikh}, {Rapisarda}, {Wijnands}, {Corbel}, {Tremou},
  {Altamirano}, {Baglio}, {Ceccobello}, {Degenaar}, {van den Eijnden},
  {Fender}, {Heywood}, {Krimm}, {Lucchini}, {Markoff}, {Russell}, {Soria}, \&
  {Woudt}}]{russell_etal_2019}
{Russell}, T.~D., {Tetarenko}, A.~J., {Miller-Jones}, J.~C.~A., {et~al.} 2019,
  \apj, 883, 198

\bibitem[{{Scepi} {et~al.}(2021){Scepi}, {Begelman}, \&
  {Dexter}}]{scepi_etal_2021}
{Scepi}, N., {Begelman}, M.~C., \& {Dexter}, J. 2021, \mnras, 502, L50

\bibitem[{{Scepi} {et~al.}(2024){Scepi}, {Dexter}, {Begelman}, {Marcel},
  {Ferreira}, \& {Petrucci}}]{scepi_etal_2024}
{Scepi}, N., {Dexter}, J., {Begelman}, M.~C., {et~al.} 2024, \aap, 692, A153

\bibitem[{{Shakura} \& {Sunyaev}(1973)}]{shakura_sunyaev_1973}
{Shakura}, N.~I. \& {Sunyaev}, R.~A. 1973, \aap, 24, 337

\bibitem[{{Shibata} \& {Tanuma}(2001)}]{shibata_tanuma_2001}
{Shibata}, K. \& {Tanuma}, S. 2001, Earth, Planets and Space, 53, 473

\bibitem[{{Singh} {et~al.}(2015){Singh}, {de Gouveia Dal Pino}, \&
  {Kadowaki}}]{singh_etal_2015}
{Singh}, C.~B., {de Gouveia Dal Pino}, E.~M., \& {Kadowaki}, L.~H.~S. 2015,
  \apjl, 799, L20

\bibitem[{{Sironi} \& {Beloborodov}(2020)}]{sironi_beloborodov_2020}
{Sironi}, L. \& {Beloborodov}, A.~M. 2020, \apj, 899, 52

\bibitem[{{Sironi} {et~al.}(2016){Sironi}, {Giannios}, \&
  {Petropoulou}}]{sironi_etal_2016}
{Sironi}, L., {Giannios}, D., \& {Petropoulou}, M. 2016, \mnras, 462, 48

\bibitem[{{Sridhar} {et~al.}(2025){Sridhar}, {Ripperda}, {Sironi}, {Davelaar},
  \& {Beloborodov}}]{sridhar_etal_2025}
{Sridhar}, N., {Ripperda}, B., {Sironi}, L., {Davelaar}, J., \& {Beloborodov},
  A.~M. 2025, \apj, 979, 199

\bibitem[{{Sridhar} {et~al.}(2021){Sridhar}, {Sironi}, \&
  {Beloborodov}}]{sridhar_etal_2021}
{Sridhar}, N., {Sironi}, L., \& {Beloborodov}, A.~M. 2021, \mnras, 507, 5625

\bibitem[{{Sridhar} {et~al.}(2023){Sridhar}, {Sironi}, \&
  {Beloborodov}}]{sridhar_etal_2023}
{Sridhar}, N., {Sironi}, L., \& {Beloborodov}, A.~M. 2023, \mnras, 518, 1301

\bibitem[{{Stella} \& {Vietri}(1998)}]{mario_luigi_1998}
{Stella}, L. \& {Vietri}, M. 1998, \apjl, 492, L59

\bibitem[{{Sturrock}(1991)}]{sturrock_1991}
{Sturrock}, P.~A. 1991, \apj, 380, 655

\bibitem[{{Tagger} {et~al.}(2004){Tagger}, {Varni{\`e}re}, {Rodriguez}, \&
  {Pellat}}]{tagger_etal_2004}
{Tagger}, M., {Varni{\`e}re}, P., {Rodriguez}, J., \& {Pellat}, R. 2004, \apj,
  607, 410

\bibitem[{{Tchekhovskoy} {et~al.}(2010){Tchekhovskoy}, {Narayan}, \&
  {McKinney}}]{tchekhovskoy_etal_2010}
{Tchekhovskoy}, A., {Narayan}, R., \& {McKinney}, J.~C. 2010, \apj, 711, 50

\bibitem[{{Tchekhovskoy} {et~al.}(2011){Tchekhovskoy}, {Narayan}, \&
  {McKinney}}]{tchekhovskoy_etal_2011}
{Tchekhovskoy}, A., {Narayan}, R., \& {McKinney}, J.~C. 2011, \mnras, 418, L79

\bibitem[{{Tohline} \& {Osterbrock}(1976)}]{tohline_osterbrock_1976}
{Tohline}, J.~E. \& {Osterbrock}, D.~E. 1976, \apjl, 210, L117

\bibitem[{{Tout} \& {Pringle}(1992)}]{tout_pringle_1992}
{Tout}, C.~A. \& {Pringle}, J.~E. 1992, \mnras, 259, 604

\bibitem[{{Tout} \& {Pringle}(1996)}]{tout_pringle_1996}
{Tout}, C.~A. \& {Pringle}, J.~E. 1996, \mnras, 281, 219

\bibitem[{{Trakhtenbrot} {et~al.}(2019){Trakhtenbrot}, {Arcavi}, {MacLeod},
  {Ricci}, {Kara}, {Graham}, {Stern}, {Harrison}, {Burke}, {Hiramatsu},
  {Hosseinzadeh}, {Howell}, {Smartt}, {Rest}, {Prieto}, {Shappee}, {Holoien},
  {Bersier}, {Filippenko}, {Brink}, {Zheng}, {Li}, {Remillard}, \&
  {Loewenstein}}]{trakhtenbrot_etal_2019}
{Trakhtenbrot}, B., {Arcavi}, I., {MacLeod}, C.~L., {et~al.} 2019, \apj, 883,
  94

\bibitem[{{Uzdensky}(2005)}]{uzdensky_2005}
{Uzdensky}, D.~A. 2005, \apj, 620, 889

\bibitem[{{Uzdensky}(2016)}]{uzdensky_2016}
{Uzdensky}, D.~A. 2016, in Astrophysics and Space Science Library, Vol. 427,
  Magnetic Reconnection: Concepts and Applications, ed. W.~{Gonzalez} \&
  E.~{Parker}, 473

\bibitem[{{Uzdensky} \& {Goodman}(2008)}]{uzdensky_goodman_2008}
{Uzdensky}, D.~A. \& {Goodman}, J. 2008, \apj, 682, 608

\bibitem[{{Uzdensky} {et~al.}(2010){Uzdensky}, {Loureiro}, \&
  {Schekochihin}}]{uzdensky_etal_2010}
{Uzdensky}, D.~A., {Loureiro}, N.~F., \& {Schekochihin}, A.~A. 2010, \prl, 105,
  235002

\bibitem[{{Werner} {et~al.}(2019){Werner}, {Philippov}, \&
  {Uzdensky}}]{werner_etal_2019}
{Werner}, G.~R., {Philippov}, A.~A., \& {Uzdensky}, D.~A. 2019, \mnras, 482,
  L60

\bibitem[{{Witzel} {et~al.}(2018){Witzel}, {Martinez}, {Hora}, {Willner},
  {Morris}, {Gammie}, {Becklin}, {Ashby}, {Baganoff}, {Carey}, {Do}, {Fazio},
  {Ghez}, {Glaccum}, {Haggard}, {Herrero-Illana}, {Ingalls}, {Narayan}, \&
  {Smith}}]{witzel_etal_2018}
{Witzel}, G., {Martinez}, G., {Hora}, J., {et~al.} 2018, \apj, 863, 15

\bibitem[{{Wood} {et~al.}(2021){Wood}, {Miller-Jones}, {Homan}, {Bright},
  {Motta}, {Fender}, {Markoff}, {Belloni}, {K{\"o}rding}, {Maitra}, {Migliari},
  {Russell}, {Russell}, {Sarazin}, {Soria}, {Tetarenko}, \&
  {Tudose}}]{wood_etal_2021}
{Wood}, C.~M., {Miller-Jones}, J.~C.~A., {Homan}, J., {et~al.} 2021, \mnras,
  505, 3393

\bibitem[{{Yang} {et~al.}(2018){Yang}, {Wu}, {Fan}, {Jiang}, {McGreer},
  {Shangguan}, {Yao}, {Wang}, {Joshi}, {Green}, {Wang}, {Feng}, {Fu}, {Yang},
  \& {Liu}}]{yang_etal_2018}
{Yang}, Q., {Wu}, X.-B., {Fan}, X., {et~al.} 2018, \apj, 862, 109

\bibitem[{{Yuan} {et~al.}(2019){Yuan}, {Spitkovsky}, {Blandford}, \&
  {Wilkins}}]{yuan_etal_2019}
{Yuan}, Y., {Spitkovsky}, A., {Blandford}, R.~D., \& {Wilkins}, D.~R. 2019,
  \mnras, 487, 4114

\bibitem[{{Zhang} \& {Giannios}(2021)}]{zhang_giannios_2021}
{Zhang}, H. \& {Giannios}, D. 2021, \mnras, 502, 1145

\end{thebibliography}

\begin{appendix}
\section{Complete Poynting Theorem analysis}
\label{sec:enbudget_full}
\begin{figure*}
    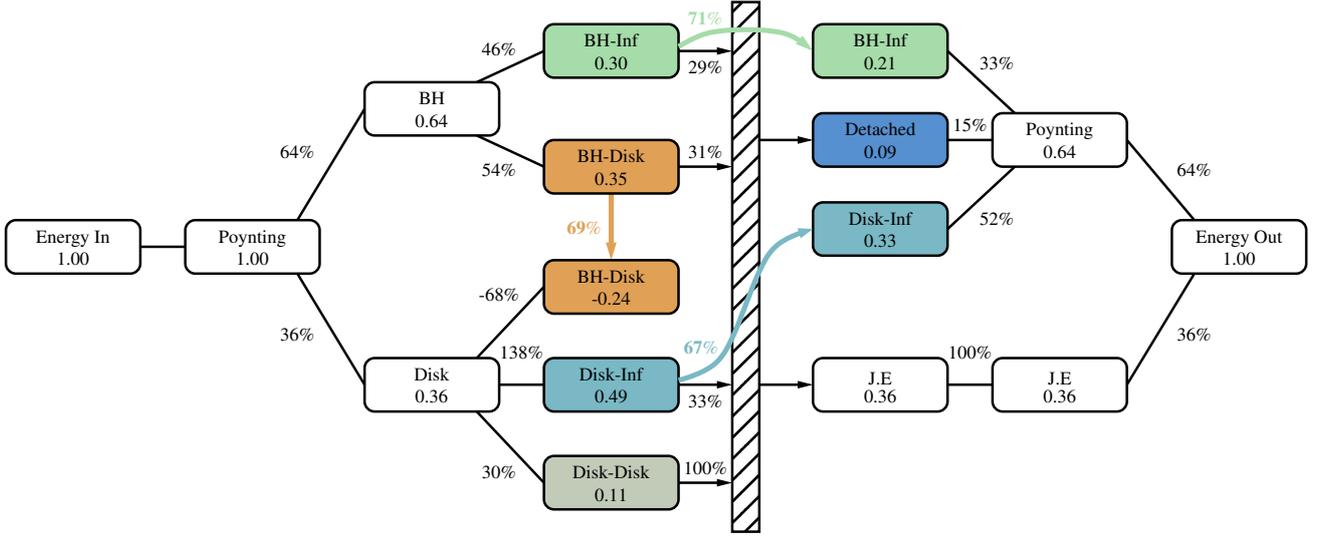
\caption{In each box, averages -- taken over~$200 \leq ct / r_g \leq 3200$ -- of contributions to Poynting's Theorem expressed as a fraction of the total input Poynting flux,~$\langle \dot{\mathcal{E}}_{\rm inj,Poynt} \rangle = 0.82 L_{\rm BZ}$ (cf.\ Table~\ref{table:edotavg}). Energy sources are on the left of the central hatched bar; energy sinks are on its right. The decomposition becomes more fine-grained toward the middle. Percentages indicate branching ratios between granularity levels. For example, from the BH-Inf box and its connection to the Poynting box on the energy sink side, one reads~$\langle \dot{\mathcal{E}}_{\rm out,BH-Inf} \rangle = 0.21 \langle \dot{\mathcal{E}}_{\rm inj,Poynt} \rangle = 0.33 \langle \dot{\mathcal{E}}_{\rm out,Poynt}$). The sum of values in every column equals unity to within percent-level error, as required for energy conservation. We exclude contributions from BH-BH field lines and the rate-of-change of the electromagnetic field energy~($\dot{\mathcal{E}}_{\rm EMfields}$) because these vanish at the percent level. Except in the cases of colored arrows crossing the hatched bar, it is not possible to determine branching ratios from an arbitrary source to an arbitrary sink. To represent this, black arrow-connections to the hatched bar indicate a sink-ambiguous deposit to (for the energy sources), or a source-ambiguous withdrawal from (for the energy sinks), the available energy budget.}
    \label{fig:poyntfull}
\end{figure*}

Here, we complement our analysis of Section~\ref{sec:enbudgetmeasure}, providing the complete decomposition of the energy budget in our simulation based on the identification of field line bundles illustrated in Fig.~\ref{fig:maglinks}. We present, in Fig.~\ref{fig:poyntfull}, the time-averaged values of every term in Poynting's theorem at hierarchical granularity levels.

As shown by Fig.~\ref{fig:poyntfull}, a good fraction of the Poynting flux injected by the BH gets directly channeled along BH-disk coupling field lines to the disk. Though the BH (BH box in the figure) injects~$64$ percent of the overall input Poynting flux into the simulation,~$24$ percent of this is immediately deposited, via BH-disk coupling field lines (BH-disk boxes in the figure) to the disk. Thus, while naively one might conclude that the BH is responsible for the majority~($64$ percent) of the simulation energy budget, in reality, of the energy that gets processed through the corona -- either reaching the edge of the box on open field lines or being dissipated as~$\vec{J} \cdot \vec{E}$ heating -- only~$64 - 24 = 40$ percent comes from the BH, while~$36+24 = 60$ percent comes from the disk.
In our companion publication \citep{crinquand_etal_inprep}, we show that for smaller loop sizes~$\lambda \lesssim \rh$, this effect becomes more extreme: a larger fraction of the magnetic flux on the BH remains closed, and so most of the energy injected by the BH is immediately absorbed by the disk. In this way, for smaller loop sizes, closed field lines connecting the BH to the disk choke jet field lines, and with them the BH's ability to transmit energy to infinity.

\end{appendix}

\end{document}